\documentclass[a4paper,fleqn,usenatbib]{mnras}

\usepackage{newtxtext,newtxmath}

\usepackage[T1]{fontenc}
\usepackage{ae,aecompl}

\usepackage{graphicx}	
\usepackage{amsmath}	
\usepackage{amssymb}	

\title[AGN feedback in low-mass galaxies]{SDSS-IV MaNGA: Evidence of the importance of AGN feedback in low-mass galaxies}

\author[S. J. Penny et al.]
{Samantha J. Penny,$^{1}$\thanks{E-mail: samantha.penny@port.ac.uk (SJP)} 
Karen L. Masters,$^{1}$
Rebecca Smethurst,$^{2}$
Robert C. Nichol,$^{1}$\newauthor
Coleman M. Krawczyk,$^{1}$
Dmitry Bizyaev,$^{3,4}$
Olivia Greene,$^{5,6}$
Charles Liu,$^{7,8,9}$\newauthor
Mariarosa Marinelli,$^{10}$
Sandro B. Rembold,$^{11,12}$
Rogemar A. Riffel,$^{11,12}$\newauthor
Gabriele da Silva Ilha,$^{11,12}$
Dominika Wylezalek,$^{13,14}$
Brett H. Andrews,$^{15}$\newauthor
Kevin Bundy,$^{16}$
Niv Drory,$^{17}$
Daniel Oravetz,$^{3}$
Kaike Pan$^{3}$
\\
$^{1}$Institute of Cosmology and Gravitation, University of Portsmouth, Burnaby Road, Portsmouth PO1 3FX, UK\\
$^{2}$School of Physics and Astronomy, University of Nottingham, University Park, Nottingham, NG7 2RD, UK\\
$^{3}$Apache Point Observatory and New Mexico State University, P.O. Box 59, Sunspot, NM, 88349-0059, USA\\
$^{4}$Sternberg Astronomical Institute, Moscow State University, Moscow, Russia\\
$^{5}$ Department of Physics and Astronomy, Vanderbilt University, 201 West End Ave, Nashville, TN 37235, USA\\
$^{6}$ Physics Department, Fisk University, 1000 17th Ave N, Nashville, TN 37208, USA\\
$^{7}$ Astrophysical Observatory, Department of Engineering Science and Physics, College of Staten Island, City University of New York,\\ 2800 Victory Boulevard, Staten Island, NY 10314, USA\\
$^{8}$ Department of Astrophysics and Hayden Planetarium, American Museum of Natural History, New York, NY 10024, USA\\
$^{9}$ Physics Program, The Graduate Center, CUNY, New York, NY 10016, USA\\
$^{10}$ School of Mathematics, Science, and Engineering, Reynolds Community College, Richmond, Virginia, 23285, USA\\
$^{11}$Departamento de Física, CCNE, Universidade Federal de Santa Maria, 97105-900, Santa Maria, RS, Brasil\\
$^{12}$Laboratório Interinstitucional de e-Astronomia, 77 Rua General José Cristino, Rio de Janeiro, 20921-400, Brasil\\
$^{13}$Center for Astrophysical Sciences, Department of Physics and Astronomy, Johns Hopkins University, 3400 North Charles Street,\\ Baltimore, MD 21218, USA \\ 
$^{14}$European Southern Observatory, Karl Schwarzschild Straße 2, 85748 Garching bei München, Germany \\
$^{15}$PITT PACC, Department of Physics and Astronomy, University of Pittsburgh, Pittsburgh, PA 15260, USA\\
$^{16}$University of California, Santa Cruz, 1156 High St. Santa Cruz, CA 95064, USA\\
$^{17}$McDonald Observatory, The University of Texas at Austin, 1 University Station, Austin, TX 78712, USA}
\date{Accepted XXX. Received YYY; in original form ZZZ}

\pubyear{2017}

\begin{document}
\label{firstpage}
\pagerange{\pageref{firstpage}--\pageref{lastpage}}
\maketitle

\begin{abstract}
We present new evidence for AGN feedback in a subset of 69 quenched low-mass galaxies ($M_{\star} \lesssim 5\times10^{9}$\,M$_{\sun}$, $M_{\rm{r}} > -19$) selected from the first two years of the SDSS-IV MaNGA survey. The majority (85\,per\,cent) of these quenched galaxies appear to reside in a group environment. We find 6 galaxies in our sample that appear to have an active AGN that is preventing on-going star-formation; this is the first time such a feedback mechanism has been observed in this mass range. Interestingly, five of these six galaxies have an ionised gas component that is kinematically offset from their stellar component, suggesting the gas is either recently accreted or outflowing. We hypothesise these six galaxies are low-mass equivalents to the ``red geysers'' observed in more massive galaxies. Of the other 62 galaxies in the sample, we find 8 do appear for have some low-level, residual star formation, or emission from hot, evolved stars. The remaining galaxies in our sample have no detectable ionised gas emission throughout their structures, consistent with them being quenched. This work shows the potential for understanding the detailed physical properties of dwarf galaxies through spatially resolved spectroscopy. 
\end{abstract}




\section{Introduction}

The role of nature versus nurture, and their respective contributions to galaxy evolution, are unclear. We know that a large fraction of high-mass galaxies in all environments have ceased star formation \citep[e.g.][]{2005Natur.433..604D,2010ApJ...709..644I,2010ApJ...721..193P}, and those massive galaxies in isolation must therefore be able to shut-off and prevent further star formation via processes internal to the galaxies themselves. It is also well known that low-mass and satellite galaxies in medium to high density environments have ceased star formation \citep[e.g.][]{2010ApJ...721..193P,2012ApJ...757...85G,2012MNRAS.419.3167S,2013MNRAS.432..336W}. One of the remaining uncertainties is the role that AGN and other intrinsic processes play in the evolution of low mass galaxies with stellar masses $M_{\star} < 5\times10^{9}$~M$_{\sun}$.

\citet{2012ApJ...757...85G} showed that the majority of quenched dwarf galaxies with stellar masses $M_{\star} < 10^{9}$~M$_{\sun}$ are found in regions of high local galaxy density, such that $>99$~per~cent of quenched dwarfs in the Sloan Digital Sky Survey \citep[SDSS,][]{2000AJ....120.1579Y} are found within a projected distance of 1.5\,Mpc and $\pm1000$\,km\,s$^{-1}$ from a galaxy with $M_{\rm{K}} < -23$. Very few quenched dwarfs are found in isolation \citep[though see][]{2017arXiv170304975J}, so environmental processes are likely required for their initial gas removal.  However, dwarf galaxies with clear AGN signatures at optical and mid-IR wavelengths are also found \citep{2013ApJ...775..116R,2014AJ....148..136M,2015MNRAS.454.3722S}. It is therefore unclear whether low-mass galaxies can be efficiently quenched or even influenced by AGN activity, or if other processes are required to shut-off their star formation. Do these AGN act to maintain quiescence in low mass galaxies, in a process similar to the red-geysers identified in \citet{2016Natur.533..504C}?

The shape of the galaxy luminosity/mass function is often explained using feedback mechanisms, such that above a characteristic mass/luminosity turnover at $M_{\star} \approx 10^{10}$\,M$_{\sun}$, AGN feedback is dominant in driving galaxy evolution and regulating star formation, while below this mass, supernova feedback or winds from high-mass stars dominate \citep[e.g.][]{2006MNRAS.370..645B,2015MNRAS.446..521S}. As a result, in many galaxy evolution models, AGN feedback is only invoked for galaxies with $M_{\star} > 10^{10}$\,M$_{\sun}$. However, \citet{2016MNRAS.463.2986S} showed that the effects of AGN feedback have the biggest impact on the stellar populations of low-to-intermediate mass galaxies with $M_{\star} < 10^{10.25}$\,M$_{\sun}$ in a sample of 1,244 Baldwin, Phillips and Terlevich  \citep[BPT][]{1981PASP...93....5B} diagram selected AGN from SDSS. Recently, \citet{2017arXiv171005900D} have proposed AGN feedback as a feedback and quenching mechanism for dwarf galaxies, which is able to drive gas out of dwarf-sized haloes more efficiently than supernova feedback. Further complicating our picture of low-mass galaxy evolution is that  a fraction of bright dwarf galaxies with $M_{\star} \sim 10^{9}$\,M$_{\sun}$ are the remnants of morphologically transformed disk galaxies, and as such, many retain evidence of this \citep[e.g.][]{2006AJ....132..497L,2014MNRAS.443.3381P,2015ApJ...799..172T,2016MNRAS.462.3955P}. This structure includes faded spiral features and embedded disks, disc-like kinematics, and may include central supermassive black holes.  

AGN are often going to be missed in wide-field searches for quenched galaxies. If a galaxy exhibits red optical colours consistent with quiescence, yet hosts strong central emission, it might be removed from a quenched galaxy sample if just the strength or equivalent width of the H$\alpha$ line is used as a diagnostic of active star formation.  In contrast, but also problematic in understanding if AGN influence the evolution of bright dwarf ellipticals, is that low mass galaxies in clusters (the most commonly studied dEs) are often gas poor objects. The central black holes in such galaxies will therefore be undetectable through emission line diagnostics, and we cannot learn anything about the role of their AGN phase in their evolution. 
 
Indeed, evidence for AGN has been found in the centre of ultra compact dwarfs \citep{2014Natur.513..398S,2017ApJ...839...72A}, thought to be the nuclear remnants of tidally stripped disk galaxies \citep[e.g.][]{2003Natur.423..519D,2013MNRAS.433.1997P}. The existence of AGN has also been identified in several dwarf-mass galaxies, through optical emission line diagnostics, X-ray observations, mid-IR diagnostics, and radio continuum surveys. For example, \citet{2015ApJ...805...12L} cross-matched 44,000 dwarf galaxies in the NASA Sloan Atlas, and found $\sim10$ galaxies with nuclear X-ray sources. \citet{2015MNRAS.454.3722S} find that just 0.7~per~cent of galaxies with $M_{\star} < 5\times10^{9}$\,M$_{\sun}$ have clear signs of hosting an AGN using the BPT diagram, \ion{He}{ii}$\lambda4686$ line, and mid-IR colours as AGN diagnostics. More recently, \citet{2017ApJ...837...66N} have found 51 low-mass galaxies with $M_{\star} < 10^{10}$\,M$_{\sun}$ that host X-ray sources, of which 37~per\,cent are centrally located within their host galaxy, a number of which also have radio counterparts. 

Low mass quenched galaxies are ubiquitous in groups and clusters, and understanding the role that AGN feedback plays in their evolution is therefore crucial for their modelling. To detect AGN in low mass galaxies, the group environment may prove more fruitful than searches in nearby clusters such as Virgo and Coma.  The weaker tidal potential of galaxy groups versus clusters, along with the lower-density intra-group medium means bright dEs in galaxy groups will likely retain more of their gas component than their cluster counterparts. Thus a wide-field Integral Field Unit (IFU) study that includes galaxies with $M_{\star} < 5\times10^{9}$\,M$_{\sun}$ is needed to fully understand the role that AGN play in the evolution of low-mass galaxies. 

In this work, we aim to identify candidate dwarf AGN hosted by low-mass galaxies in the SDSS-IV MaNGA survey, which includes low-mass galaxies in all environments with $M_{\star} < 5\times10^{9}$\,M$_{\sun}$. In particular, we aim to identify red geyser candidates \citep[][]{2016Natur.533..504C} in the low-mass galaxy sample, which typically have kinematically offset ionised gas, either the result of accretion or outflows of material generated by AGN-driven winds. Spatial data is required to map these stellar and ionised gas velocity components, as well as identify the bi-symmetric emission line features seen in red geysers which would not be identified in single-fibre or long-slit spectroscopic surveys. 

We present an overview of the MaNGA observations used for this work in Section\,\ref{sec:obs}, with the sample selection described in Section\,\ref{sec:whanclass}. We examine the ionised gas emission line mechanism in Section\,\ref{sec:iongas}, with spaxel-by-spaxel emission line diagnostics presented  in Section\,\ref{sec:bpt}, and mid-IR colours for the quenched, low-mass galaxies examined in Section\,\ref{sec:midir}. We discuss our results in Section\,\ref{sec:discuss}, and conclude in Section\,\ref{sec:conclude}. Throughout this paper, we assume a $\Lambda$CDM cosmology, with $H_{0} = 70$~km~s$^{-1}$, $\Omega_{M} = 0.3$, and $\Omega_{\Lambda} = 0.7$. All magnitudes are presented in the AB magnitude system, unless otherwise stated. 

\section{Observations and Sample Selection}
\label{sec:obs}

\subsection{The MaNGA Survey}
\label{sec:manga}

The data for this work is drawn from the Sloan Digital Sky Survey IV \citep[SDSS-IV,][]{2017arXiv170300052B} Mapping Nearby Galaxies at APO \citep[MaNGA,][]{2015ApJ...798....7B} survey. Searching for the processes responsible for quenching requires high quality, spatially resolved spectroscopic data, in order to trace key spectral features that will reveal the quenching mechanisms active in a given galaxy. MaNGA is a multi-object IFU survey using the Baryon Oscillation Spectroscopic Survey (BOSS) spectrograph \citep{2013AJ....146...32S} on the 2.5\,m Sloan Foundation Telescope \citep{2006AJ....131.2332G}. MaNGA will target 10,000 galaxies with $M_{\star} \gtrsim 10^{9}$~M$_{\sun}$ by the completion of the survey in 2020, which began taking data on 2014 July 1. The MaNGA targets are selected independent of colour,  morphology and environment, ensuring we draw targets from a range of local galaxy density, crucial in a study of the processes responsible for galaxy quenching. 

MaNGA has a spectral resolution $R\sim2000$, which corresponds to an instrumental resolution $\sim70$\,km\,s$^{-1}$ in the vicinity of the H$\alpha$ feature. The data span a large wavelength range of 3600\,\AA\ to 10300\,\textrm{\AA} which covers key spectral features from the optical to the near-IR crucial for measuring stellar populations and extracting both stellar and gas kinematics. Each galaxy in the survey is targeted with one of 17 IFU fibre bundles per plate, with the bundles containing 19 ($\times2$), 37 ($\times4$), 61 ($\times4$), 91 ($\times2$), or 127 ($\times5$) fibres \citep[see][ for further details]{2015AJ....149...77D}. The size of the fibre bundle typically matched to the angular size of the galaxy, such that the majority of the targets have spectral coverage to at least 1.5\,$R_{\rm{e}}$. The fibres within the bundles have radius $2$\,arcsec, which matches the typical ground-based seeing at APO. During the observations, a three-point dither pattern is applied, such that the inter-fibre gaps in each IFU bundle are sampled \citep{2015AJ....150...19L}. The final reduced datacubes are resampled to square spaxels of size $0.5$\,arcsec with a median spatial resolution of $\sim2.5$\,arcsec FWHM \citep{2016AJ....152...83L}.
The flux calibration of the data is better than 5\,per\,cent \citep{2016AJ....151....8Y} for 89\,per\,cent of the wavelength range, and 1.7\,per\,cent between H$\beta$ and H$\alpha$. 

MaNGA has three main samples, the primary sample, the secondary sample, and the colour-enhanced sample, which are selected to meet several different science goals. The primary sample is designed such that $\sim80$\,per cent of the galaxies in this sample have IFU coverage to at least 1.5\,$R_{\rm{e}}$, and the sample has a mean redshift $\langle z \rangle = 0.03$. 
The primary sample also includes a ``colour-enhanced'' sample, which contains galaxies selected to lie in underrepresented regions of the colour magnitude diagram, including high-mass blue galaxies, ``green valley'' galaxies, and, importantly for this study, low-mass red galaxies. These colour-enhanced galaxies are also targeted to 1.5\,$R_{\rm{e}}$. The secondary sample is designed such that  $\sim80$\,per cent of the galaxies have IFU coverage to at least 2.5\,$R_{\rm{e}}$, and as such, has a higher mean redshift than the primary sample. For further details of the survey design, see \citet{2016AJ....152..197Y}.

\subsubsection{Mid-Infrared Photometry}

We also use mid-infrared (IR) photometry in this work, which is not provided as part of the NASA Sloan Atlas. The mid-IR photometry used in this work is taken from the \textit{Wide Field Infrared Survey Explorer} \citep[\textit{WISE}][]{2010AJ....140.1868W} All Sky Survey. \textit{WISE} is a mid-IR space based telescope, which imaged the whole sky in four photometric bands: 3.4~$\mu$m (W1), 4.6~$\mu$m (W2), 12~$\mu$m (W3), and 22~$\mu$m. The 3.4~$\mu$m and 4.6~$\mu$m bands are sensitive to stellar light and hot dust respectively, making $[3.4]-[4.6]$ colours ideal for the identification of certain types of AGN. The 12~$\mu$m band traces crucial lines associated with recent star formation (e.g. and the line), and can thus be used to separate truly quenched galaxies from those with star formation in the past 2\,Gyr. We matched the positions of the MaNGA low-mass galaxy sample to the \textit{WISE} All Sky Survey photometry catalogue, and photometry was found for 303/310 of the low-mass galaxies we selected in Section\,\ref{sec:whanclass}. The \textit{WISE} photometry is left in its native Vega magnitude system. 

\subsection{Separating star forming and quenched galaxies}
\label{sec:whanclass}

The MaNGA survey targets galaxies with $10^{9}\,\textrm{M$_{\sun}$} \lesssim M_{\star} \lesssim 5\times10^{11}$\,M$_{\sun}$, and we select the lowest mass galaxies in the survey for this work. This builds on the sample of low-mass, quenched galaxies presented in \citet{2016MNRAS.462.3955P}, which was selected from galaxies in SDSS Data Release 13 (DR13, Albereti et al., submitted). We first identify all low-mass, dwarf-like galaxies in SDSS-MaNGA with spectra reduced prior to June 2016 (known internally as MaNGA MPL-5), which were processed using Version 2.0.1 of the MaNGA Data Reduction Pipeline. This sample was released in SDSS data release 14 \citep[DR14][]{2017arXiv170709322A}. We identify dwarf galaxies as all galaxies with absolute magnitudes fainter than $M_{\rm{r}} \approx -19$, central stellar velocity dispersions $\sigma_{\star} < 100$~km~s$^{-1}$, and stellar mass $M_{\star} < 5\times10^{9}$\,M$_{\sun}$ ($M_{\star} \lesssim 2.5\times10^{9}$\,M$_{\sun}$ as listed in the NASA Sloan Atlas, which assume $h = 1$). 310 MaNGA galaxies meet these criteria. The galaxies in this sample have $ -19.04 < M_{\rm {r}}  < -15.85$,  and colours $0.64<(u-r)<3.59$. The median stellar mass of the sample is $2.3 \times10^{9}$\,M$_{\sun}$ assuming $h=0.7$, comparable with that of the Large Magellanic Cloud \citep[$2.3\times10^{9}$\,M$_{\sun}$][]{2011MNRAS.411..495J}. This is similar to the selection criteria used in \citet{2016MNRAS.462.3955P}. 

All spectra within 1~$R_{\rm{e}}$ are stacked to provide a high signal-to-noise spectrum for each galaxy in our sample, and are provided as part of Version 2.0.2 of the MaNGA data analysis pipeline (DAP, Westfall et al., in prep). These binned spectra provide an ideal ``first look'' at the galaxy, allowing for the quick identification of the integrated properties of an individual galaxy, without having to examine the galaxy spaxel-by-spaxel.  This combined spectrum will include any star forming regions within 1~$R_{\rm{e}}$, which will be identified through the presence of emission in the stacked spectrum. A single object with extremely low $S/N = 2$ is removed, as the resulting spectrum is unusable for our science analysis, with the remaining stacked spectra having $11 < S/N < 180$. 

Emission-line strengths are provided as part of the MaNGA data analysis pipeline. First, as part of the DAP fit, the emission lines are masked, and the stellar continuum is modelled using the kinematic and stellar population fitting package p\textsc{pxf} \citep{2004PASP..116..138C}. The stellar continuum model is constructed using a thinned version of the MILES spectral library, and due to the wavelength coverage of the library, the model spans the wavelength range $ 3525\,\textrm{$\AA$} < \lambda < 7500$\,\textrm{$\AA$}.  The model is also broadened to match the stellar velocity dispersion of the galaxy, and thus the absorption lines can be cleanly subtracted from the spectrum. The residual emission lines are then modelled using Gaussian profiles, with 21 different lines fit in total. The summed line fluxes, equivalent widths, and velocity dispersions are provided as part of the DAP output. 

To build a sample of low-mass quenched galaxies, we must first remove any star-forming dwarfs using line strength diagnostics. 
We identify such galaxies using emission line-strength diagnostic methods, as colour alone cannot be used to accurately characterise a galaxy as quenched or star forming, as dusty star-forming galaxies will remain in the sample. 
However, no single line-strength diagnostic system can reliably class all galaxies as star forming, quenched, or hosting AGN. 
 For example, if the spectrum of a galaxy does not exhibit the H$\beta$ feature in emission due to dust obscuration or weak emission, it cannot be placed on the \citep[BPT][]{2008MNRAS.391L..29S} diagram.
 
 \subsubsection{WHAN classification}
 
To identify quenched, low-mass galaxy candidates from our sample of dwarf galaxies, we therefore use the width of H$\alpha$ vs. [\ion{N}{ii}] line strength diagnostic \citep[WHAN,][]{2011MNRAS.413.1687C}. This diagram uses the line strength ratio of the two most prominent emission lines in most galaxy spectra: the H$\alpha$ and [\ion{N}{ii}]$\lambda6583$ lines. This classification scheme uses the equivalent width of the H$\alpha$ line as a proxy for the ratio of the intrinsic H$\alpha$ luminosity to the H$\alpha$ luminosity expected due to stellar populations older than $10^{8}$\,yr.  This allows for the separation of emission due to star formation and AGN, from that originating from hot, evolved stars. The flux ratio between [\ion{N}{ii}] and H$\alpha$ is used to separate AGN activity from star formation for galaxies with EW$_{\rm{H\alpha}} > 3$\,\textrm{\AA}. The WHAN diagram also allows for the identification of galaxies in which the heating of their ionised gas is the result of old stars, rather than star formation or AGN activity, as well as those galaxies that cannot be classified via the BPT diagram due to absent absorption lines. For full details of the WHAN classification scheme, see \citet{2011MNRAS.413.1687C}. 

First, star forming galaxies are removed from the sample. These galaxies have log([\ion{N}{ii}]/H$\alpha) < -0.4$ and EW$_{\rm{H\alpha}} > 3$\,\AA\ under the WHAN classification system, and 217 galaxies meet these criteria. 93 galaxies remain after the removal of star forming galaxies from our sample.  

We also identify all emission-line free galaxies using the WHAN line-strength diagnostic method. These galaxies are identified by the absence of significant H$\alpha$  and [\ion{N}{ii}] emission lines in their spectra, with both lines required to have equivalent widths $<0.5$\,\textrm{\AA}, else no detectable H$\alpha$ emission. 57 galaxies meet these criteria. We then identify all galaxies with weak H$\alpha$ and [\ion{N}{ii}] lines, which are inconsistent with AGN activity- the ``retired'' galaxy populations in the \citet{2011MNRAS.413.1687C} WHAN classification system. Their emission lines are the result of heating by hot, old, evolved, low-mass stars, and such objects have $0.5 < EW_{\rm{H\alpha}} < 3$. Fourteen low mass galaxies in our sample are ``retired'' objects, with emission line strengths inconsistent with ongoing star formation. The WHAN classification system therefore identifies 71 low-luminosity galaxies in MaNGA MPL5 that can be considered to be passively evolving. 

Those galaxies classified as quenched with clear star formation or blue spiral morphologies are manually removed from the quenched galaxy sample.  A visual inspection of its central spectrum revealed clear emission lines that were not properly measured by the DAP. This process removed 1 object from the quenched galaxy sample. We also remove one object at $z=0.278$ which had made our low-mass quenched galaxy sample. The NASA Sloan Atlas listed the redshift  of this galaxy as $z=0.0286$, and the galaxy's absolute photometry was therefore incorrect. This process left 69 candidate galaxies that host little or no star formation. 33 objects in our quenched low-mass galaxy sample are drawn from the primary sample, 4 from the secondary sample, and 32 from the colour-enhanced sample.  

The remaining 22 galaxies are classified as AGN on the WHAN diagram. The ten objects with EW$_{\rm{H\alpha}} < 6$\,\AA\ are weak AGN, in which the emission lines are likely the combination of both AGN and star formation activity. The remaining twelve galaxies have EW$_{\rm{H\alpha}} > 6$\,\AA\ classified as Seyfert-like emission. However, we note the majority of these are edge-on disk galaxies, and are likely heavily dust obscured. Nevertheless, we go on to examine the their spaxel-by-spaxel BPT diagrams. 

\begin{figure}
\includegraphics[width=\columnwidth]{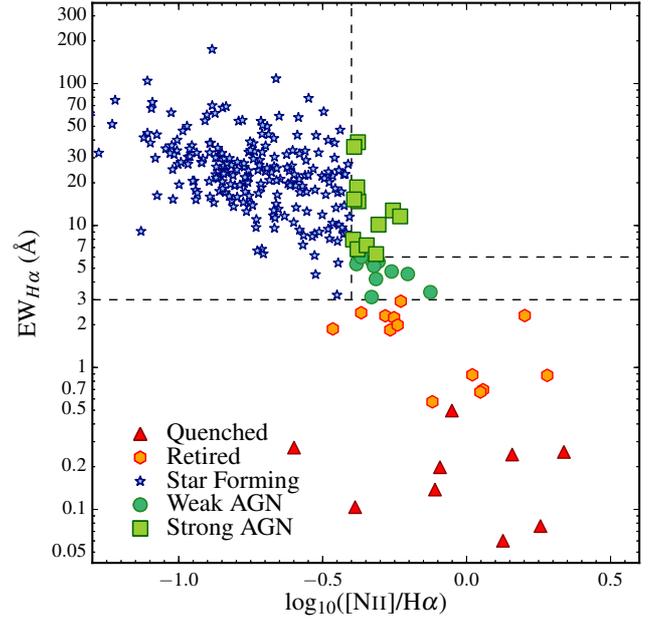}
\caption{Equivalent width of H$\alpha$ versus [\ion{N}{ii}]/H$\alpha$ \citep[WHAN][]{2011MNRAS.413.1687C} diagram for faint ($M_{\rm{r}} > -19$)  galaxies in the MaNGA survey. The dashed lines are the diagnostic lines from, which split the diagram into regions where the emission is dominated by star formation, AGN, or heating by hot, old stars. The majority of the galaxies (217/310) below the luminosity cut of $M_{\rm{r}} > -19$ are star forming systems.}
\label{fig:whan}
\end{figure}

\subsection{Local environment of quenched versus star forming low-mass galaxies}
\label{sec:lenv}

\begin{figure}
\includegraphics[width=\columnwidth]{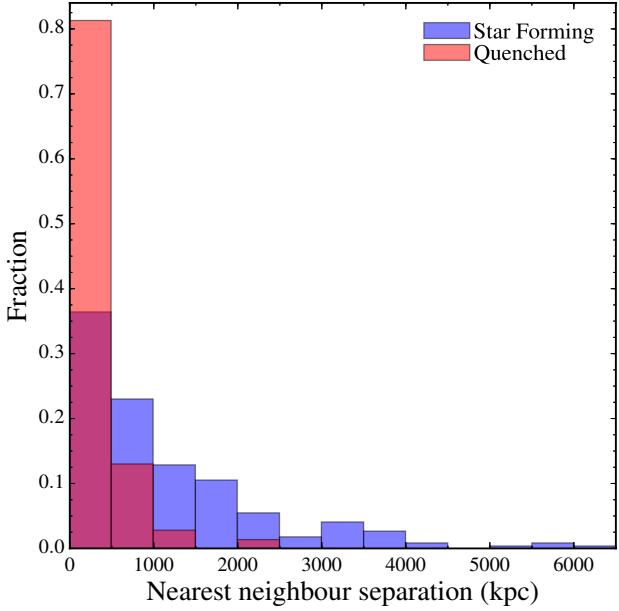}
\caption{Histogram showing the separation of the low-mass galaxies from their nearest bright neighbour galaxy with $M_{\rm{K}} < -23$. The distribution for both the quenched and star forming low-mass galaxy samples as selected in Section\,\ref{sec:whanclass} are shown. The distribution of the quenched galaxies are shown as red bars, and the star forming galaxies by blue bars. The purple regions are the overlap between the star forming and quenched objects. }
\label{fig:env}
\end{figure}

We quantify the local environment of the low-mass quenched galaxies using both a  neighbour method and a local galaxy density measure. First, we trace large-scale structure using a combination of the 2MASS redshift survey and SDSS. This redshift catalogue is complete to $M_{\rm{K}} = -23$ at $z=0.04$, comparable to the highest redshift of our MaNGA low-mass galaxy sample. Identical to \citep{2016MNRAS.462.3955P}, we find the  bright neighbour galaxy with $M_{\rm{K}} < -23$ within $\pm1000$\,km\,s$^{-1}$ for each low-mass galaxy in our sample. We also count the number of bright galaxies with $M_{\rm{K}} <-23$ within a radius of 1.5\,Mpc (comparable to the virial radius of a galaxy cluster) and $\pm1000$\,km\,s$^{-1}$ of each low-mass galaxy. 

The  neighbour distances of our quenched and star forming samples as defined in Section\,\ref{sec:whanclass} are shown in Fig.\,\ref{fig:env}. For clarity, the WHAN-classified AGN are not included on this plot. A clear difference in environment is seen for the two samples, such that quenched galaxies are typically found at separations $<1000$\,kpc from their  bright neighbour galaxy. The low-mass, quenched galaxies have separations $17~\textrm{kpc} < D_{bright} < 4582$\,kpc from their  bright neighbour, with a median separation $228\pm72.8$\,kpc. In contrast, the purely star-forming objects have separations $32~\textrm{kpc} < D_{bright} < 6056$\,kpc, with a median separation $706\pm78.5$\,kpc. A difference in local galaxy density is also seen, such that within a comoving radius of 1.5\,Mpc and a velocity separation $\pm1000$\,km\,s$^{-1}$, a quenched low-mass galaxy has a mean of $16\pm2.8$ bright neighbours with $M_{\rm{K}} < -23$, whereas a star forming low-mass galaxy has a mean $2.2\pm2.55$ bright neighbours using an identical magnitude limit. 

\section{Ionised gas in quenched galaxies}
\label{sec:iongas}

We search for any objects in our low-mass, quenched galaxy sample that retain an ionised gas component. If a galaxy retains a gas reservoir, or if it is actively accreting gas from its surroundings, yet is no longer forming stars, then a feedback mechanism must be operational within that galaxy.  By identifying those quenched galaxies that exhibit ionised gas, and quantifying their emission line flux ratios and the location within the galaxy of the emission, we can determine the heating mechanism acting upon the gaseous component. 

\begin{figure*}
\centering
\includegraphics[width=0.78\textwidth]{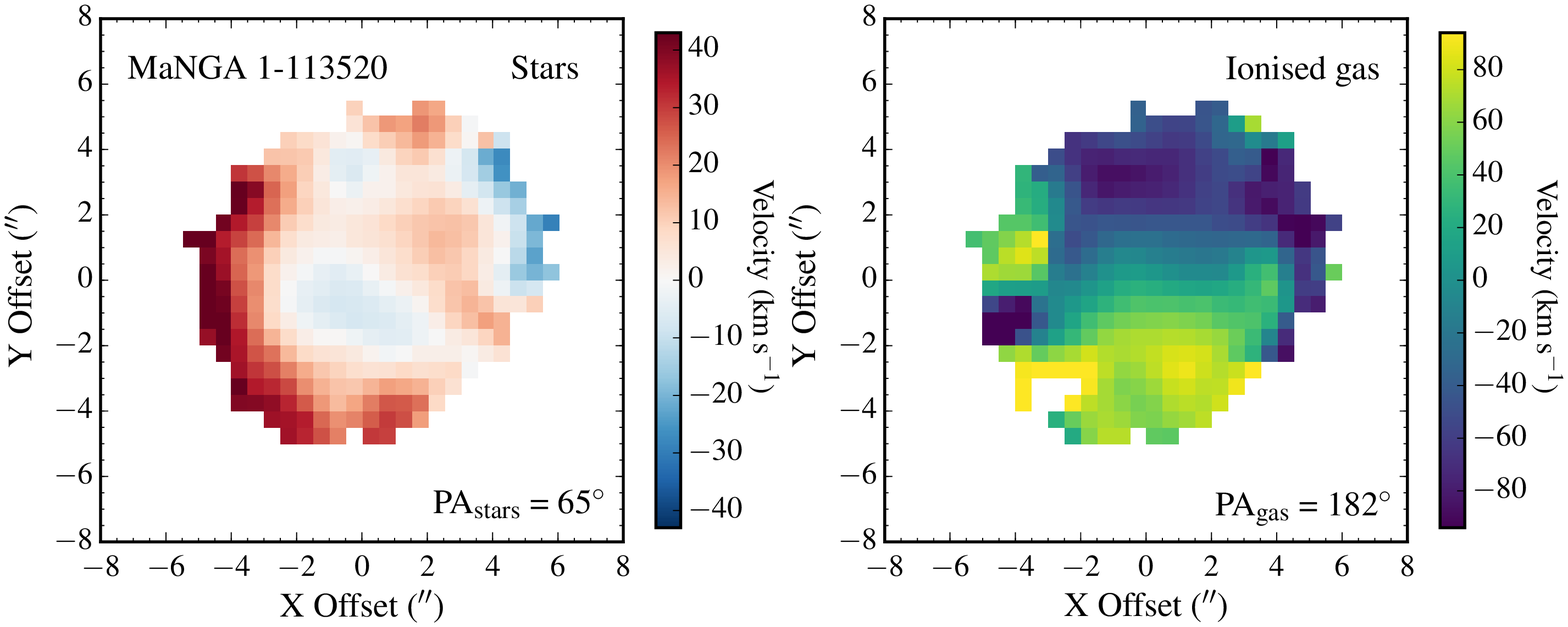}\\
\includegraphics[width=0.78\textwidth]{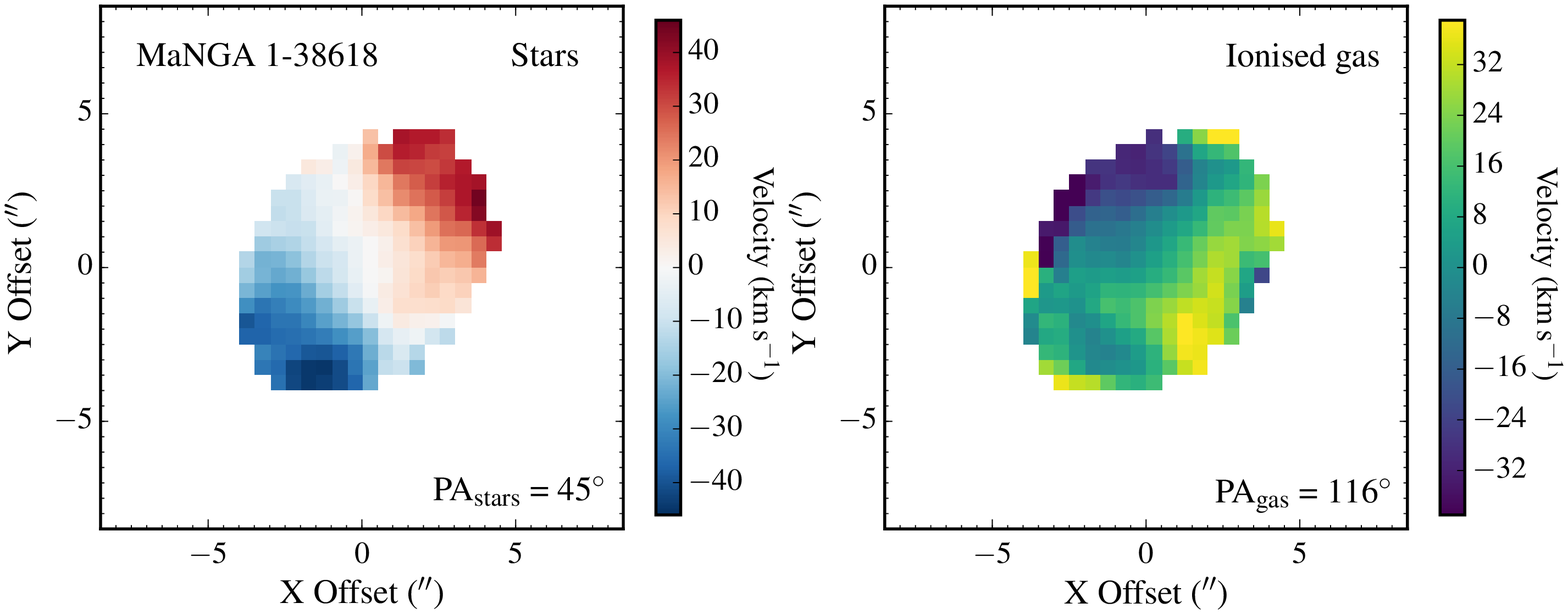}\\
\includegraphics[width=0.78\textwidth]{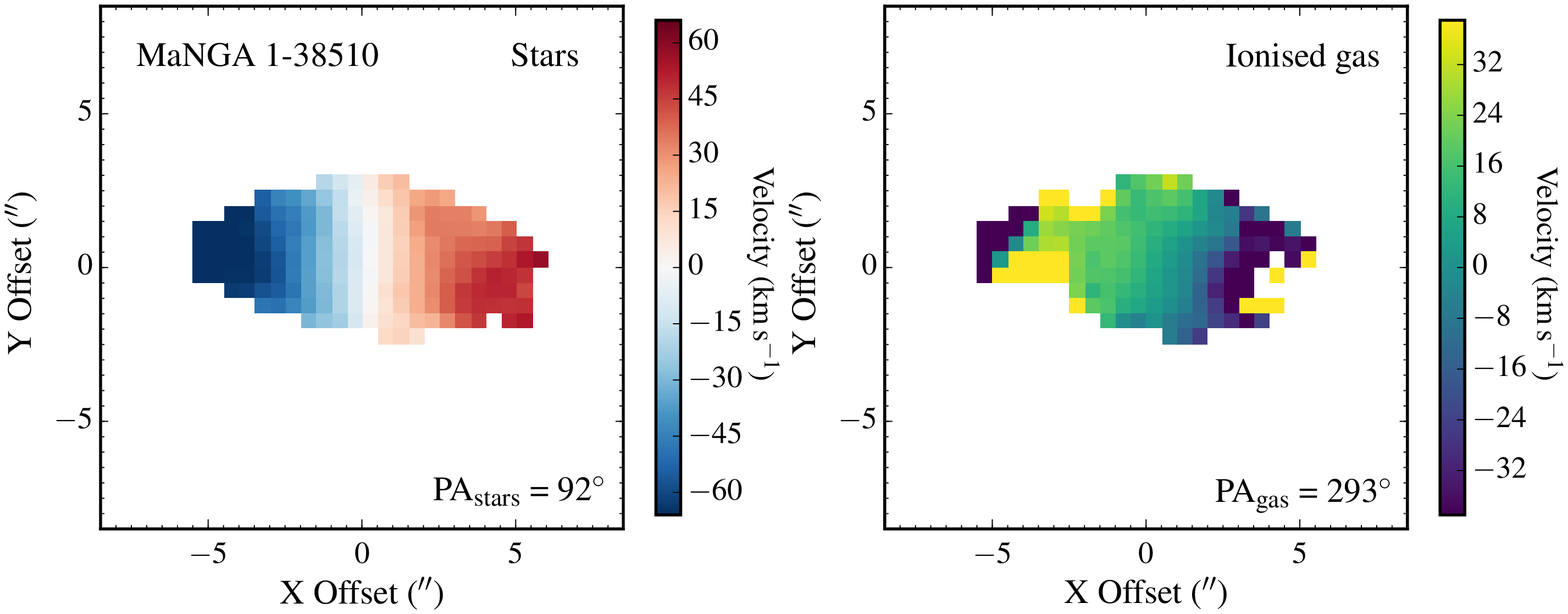}\\
\includegraphics[width=0.78\textwidth]{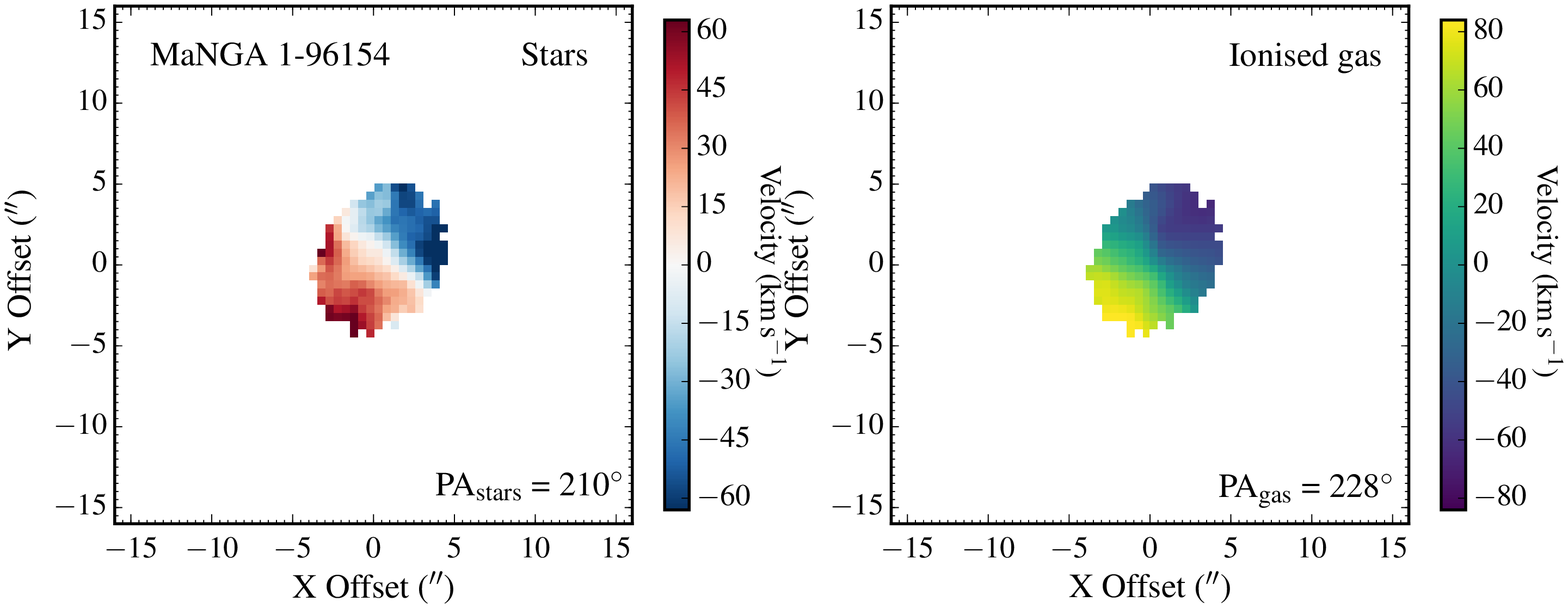}\\
\caption{Example stellar velocity maps (left panel) and ionised gas velocity maps (right panel) for 4/14 MaNGA galaxies in our sample which contain an ionised gas component. We trace the ionised gas velocity feature using the H$\alpha$ line. Two galaxies with co- or counter-rotating stellar and ionised gas components are shown , along with two galaxies that host kinematically offset stars and ionised gas. All have $M_{\rm{r}} > -19$, $M_{\star} \lesssim 5\times10^{9}$\,M$_{\sun}$, and $\sigma_{\star} <100 $~km\,s$^{-1}$, placing them in the dwarf galaxy regime. The remaining velocity maps are shown in Appendix\,\ref{appendix:vmaps}.}
\label{fig:vfields}
\end{figure*}

Gas that is either co-rotating with the galaxy's stellar component, else offset by exactly 180$\degr$, has both components in dynamical equilibrium. However, when rotation of the ionised gas component is randomly offset, this is likely evidence of an accretion event. First, we identify all low mass galaxies with clear rotation in their [\ion{O}{III}]$\lambda5008$, H$\alpha$, or [\ion{N}{ii}]$\lambda6583$ emission line features. These three lines are used as they are very sensitive tracers of ionisation, and typically the most prominent emission lines in a galaxy spectrum. The ratio between these features can be used to separate AGN heating from both star formation, and heating by old stellar populations (e.g. Cid-Fernandes et al., 2011). From our initial sample of 63 quenched dwarf galaxies, 12 are identified with clear rotation in both their stellar and ionised gas components. One object, MaNGA\,1-38166,  has a rotating ionised gas component, but no coherent stellar rotation. Excluding a single member of the Coma Cluster (MaNGA 1-456355), the quenched low-mass galaxies that retain an ionised gas component are found in galaxy groups with fewer than seven $L_{\star}$ or brighter galaxies (see Section\,\ref{sec:lenv}). Velocity maps showing the stellar and ionised gas velocity fields for each low-mass galaxy with clear rotation are shown in Fig.\,\ref{fig:vfields}. 

As can be easily seen in Fig\,\ref{fig:vfields} and Appendix\,\ref{appendix:vmaps}, the stars and ionised gas are not necessarily co-rotating, and thus not all of the galaxies are in dynamical equilibrium. We therefore go on to compare the kinematic position angles of the stellar and ionised gas components of these galaxies. The global kinematic position angles of the two components are measured using the method described in Appendix C of \citet{2006MNRAS.366..787K}. We note here that a number of the offset ionised gas velocity fields may be outflows (e.g. MaNGA\,1-113520), so we use this measure only to check for gas that is neither co- or counter-rotating with the stellar component. 

The average velocities of the H$\alpha$ and [\ion{N}{ii}]\textrm{$\lambda$}6583 lines are used in the calculation of the ionised gas kinematic PA, as they are typically among the strongest emission lines in the galaxy spectra. The velocities of the individual emission lines are output by the MaNGA-DAP, and are calculated using single-profile gaussian fits to the emission lines after the stellar continuum has been subtracted. The global kinematic PAs of the H$\alpha$ and [\ion{N}{ii}]\textrm{$\lambda$}6583 velocity fields typically agree to $<20\degr$ for all objects. We set a range of $0\degr$ to 360$\degr$ for the limits on the kinematic position angles to identify galaxies with counter-rotation between their stars  and ionised gas. 

Prior to the determination of the kinematic position angles, the stellar and ionised gas velocity maps are convolved with a gaussian filter to remove any noise which could affect the position angle measurements, especially for the galaxies targeted with the 19 fibre bundles. The filter size is chosen to be 2 spaxels, to remove artefacts on the scale of the noise from the velocity maps. For MaNGA\,1-113520, which contains a counter-rotating core in its stellar component, we do not attempt to fit a single global kinematic position angle. Instead, we only fit the kinematic position angle for the centrally rotating component, with the outer component assigned a kinematic position angle offset by $180\degr$ from the central region. We were unable to determine a stellar kinematic PA for MaNGA\,1-38166, though a visual inspection of its stellar kinematic map suggests it hosts weak stellar rotation. The measured angles for both the stellar and gaseous components are given in Table\,\ref{tab:kinpas}, along with their central stellar velocity dispersions and S\'ersic indices. 

\begin{figure}
\centering
\includegraphics[width=\columnwidth]{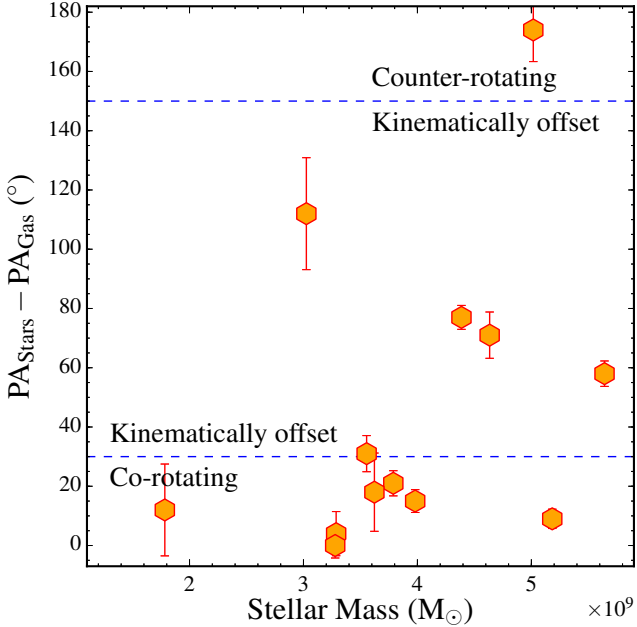}
\caption{Difference between the stellar and the ionised gas kinematic position angles as a function of galaxy mass. Object with kinematic offsets $>180\degr$ between their stellar and ionised gas components have been subtracted by $180\degr$ for simplicity. No obvious trend for galaxies to host kinematically offset gas with stellar mass is seen. One object, MaNGA\,1-38166 could not be included on this plot as we could not reliably measure a stellar kinematic position angle.}
\label{masskins}
\end{figure}

For eight galaxies in our sample, the recovered kinematic position angles of both their stars and ionised gas have difference $\Delta$PA\,$<30\degr$ or $\Delta$PA\,$>150\degr$. These small kinematic offsets show the stars and gas within these objects are likely in dynamical equilibrium. However, clear differences between the two kinematic position angles are seen for the remaining five galaxies, and these objects likely host recently accreted gas or satellites. A stellar kinematic position angle could not be determined for MaNGA\,1-38166, though a visual inspection of the map suggests its stars and gas are co-rotating. In Fig.\,\ref{masskins}, we plot $\Delta$PA as a function of galaxy stellar mass, and there is no preference for galaxies hosting kinematically offset gas to either higher or lower stellar masses. To characterise the heating mechanism of this ionised gas, we go on to examine the nature of their emission on the BPT diagram.

\begin{table*}
\begin{center}
\caption{Basic properties for the 14 low-mass galaxies identified as quenched galaxies in Section\,\ref{sec:iongas} which host an ionised gas component. Global kinematic position angles for their stellar (PA$_{\star}$) and ionised gas  (PA$_{\rm{gas}}$) components, along with the difference between the two values $\Delta$PA are provided. A global stellar kinematic PA could not be reliably measured for MaNGA 1-113520 due to the presence of a counter-rotating component.}
\begin{tabular}{lccccccccccc}
\hline
  MaNGA-ID & Plate & IFU  & RA & Dec & $z$  & $M_{\rm{r}}$  & S\'ersic $n$&  $\sigma_{\star}$  & PA$_{\rm{stars}}$ & PA$_{\rm{gas}}$ & $\Delta$PA \\
   & & & (J2000.0) & (J2000.0) & &  (mag) &  & (km\,s$^{-1}$) & ($\degr$) & ($\degr$) &  ($\degr$) \\
\hline
  1-38166    & 8081 & 3702 & 03:19:47.24 & +00:37:25.8 & 0.025 & -18.65 & 3.4 &  \ldots   & $\ldots$  & $86 \pm 7.6$ & $\ldots$\\   
  1-38618    & 8084 & 1902 & 03:30:29.42 & -00:29:19.6 & 0.022 & -18.84 & 2.7 & 43          &$ 45  \pm   6.5$ & $116 \pm 9.4$ & 71\\
  1-38510    & 8155 & 1901 & 03:32:05.59 & +00:28:45.8 & 0.022 & -18.73 & 1.6 &  \ldots    & $92   \pm 1.5$  & $113 \pm 7.9$ & 21\\ 
  1-379255  & 8711 & 1901 & 07:53:03.98 & +52:44:35.5 & 0.018 & -18.36 & 4.7 & 78          & $134\pm18.4$ & $22 \pm 9.2$ & 112\\    
  1-230177  & 8942 & 6101 & 08:19:35.49 & +26:21:45.6 & 0.020 & -18.82 & 1.8 & \ldots     & $37 \pm 2.2$ & $52 \pm 5.0$ & 15\\  
  1-488575  & 8449 & 1902 & 11:13:49.70 & +22:48:38.4 & 0.022 & -18.46 & 2.2 & \ldots     & $67   \pm  6.6$   & $71 \pm 6.0$ & 4\\ 
  1-456355  & 8931 & 6104 & 12:57:11.96 & +27:06:12.0 & 0.025 & -19.02 & 1.6 & \ldots     & $167 \pm 2.7$ & $176 \pm 1.7$ & 9\\    
  1-284335  & 8318 & 6103 & 13:07:17.01 & +45:43:41.3 & 0.035 & -19.04 & 4.0 & \ldots     & $207   \pm  9.7$ & $21 \pm 9.9$ & 186\\
  1-93551    & 8483 & 1901 & 16:25:14.66 & +48:43:16.8 & 0.021 & -17.86 & 6.0 & 49          & $101 \pm 14.9$ & $113 \pm 9.1$ & 12\\ 
  1-96154    & 8612 & 9102 & 16:57:35.35 & +39:17:09.9 & 0.033 & -18.87 & 1.0 & \ldots      & $210 \pm 12.1$ & $228 \pm 14.0$ & 18\\ 
  1-136305  & 8606 & 3704 & 17:03:39.73 & +36:23:05.8 & 0.025 & -18.45 & 1.1 & 56          & $278 \pm 2.6$ & $278 \pm 5.5$ & 0\\    
  1-178823  & 8623 & 9102 & 20:47:03.31 & +00:26:12.4 & 0.013 & -18.92 & 2.0 & \ldots      & $97   \pm  1.7$ & $354 \pm 6.7$ & 257\\
  1-113520* & 7815 & 1901 & 21:10:00.53 & +11:30:38.3 & 0.017 & -18.98 & 2.3 & 52         & $65 \pm 4.5^{*}$ & $182 \pm 1.8$ & 58\\   
  1-29809    & 8655 & 1902 & 23:53:52.52 & -00:05:55.4 & 0.022 & -18.72 & 1.8 & \ldots        &$ 21  \pm  4.4$ & $232 \pm 8.9$ & 211\\  
\hline\end{tabular}
\label{tab:kinpas}
\end{center}
\begin{flushleft} 
\smallskip $^{*}$MaNGA 1-113520 hosts a counter-rotating core, and thus the measured kinematic position angle is for the core region only. 
\end{flushleft}
\end{table*}

\subsection{Spaxel-by-spaxel BPT diagrams}
\label{sec:bpt}

\begin{figure*}
\centering
\includegraphics[width=0.45\textwidth]{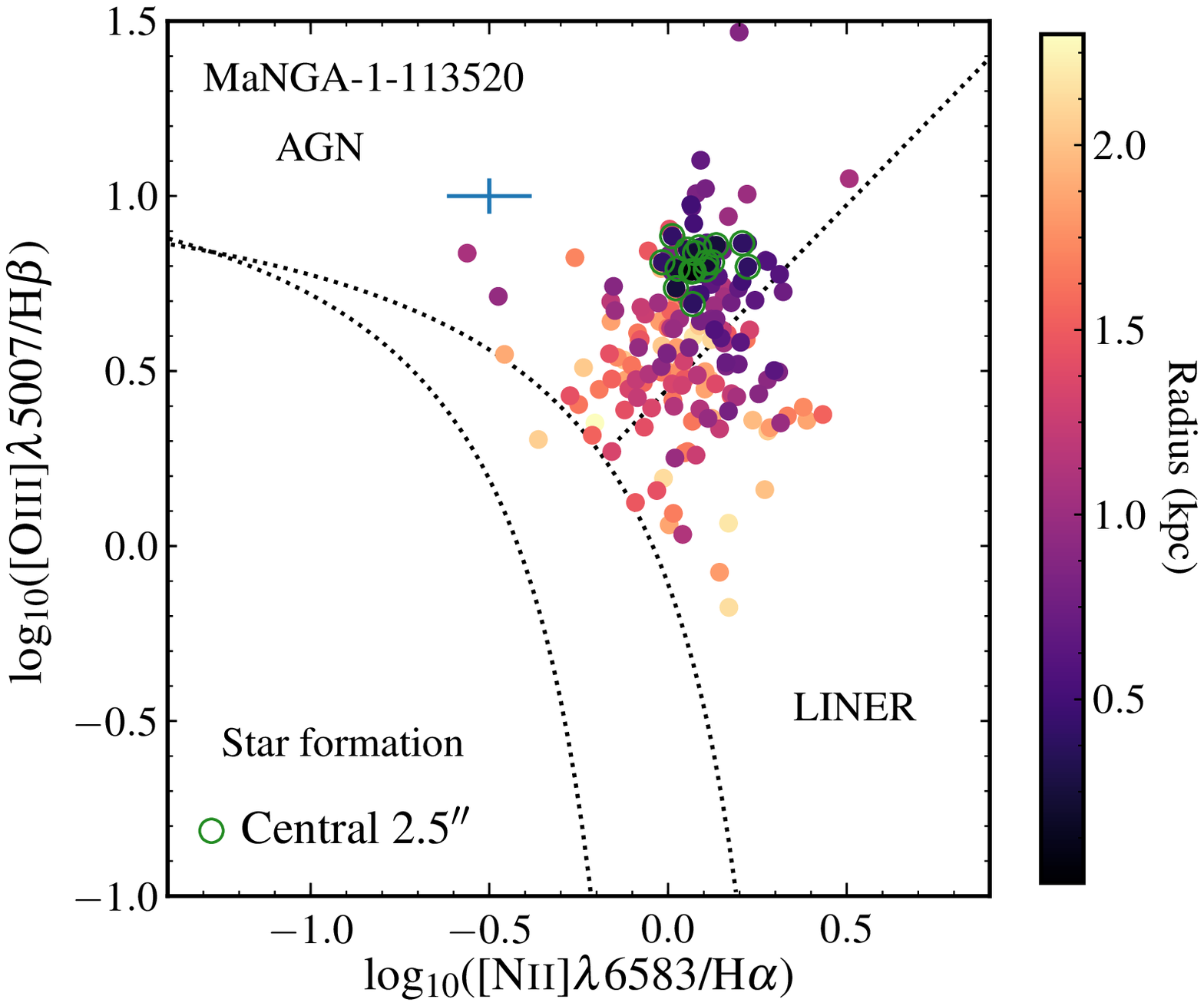}
\includegraphics[width=0.45\textwidth]{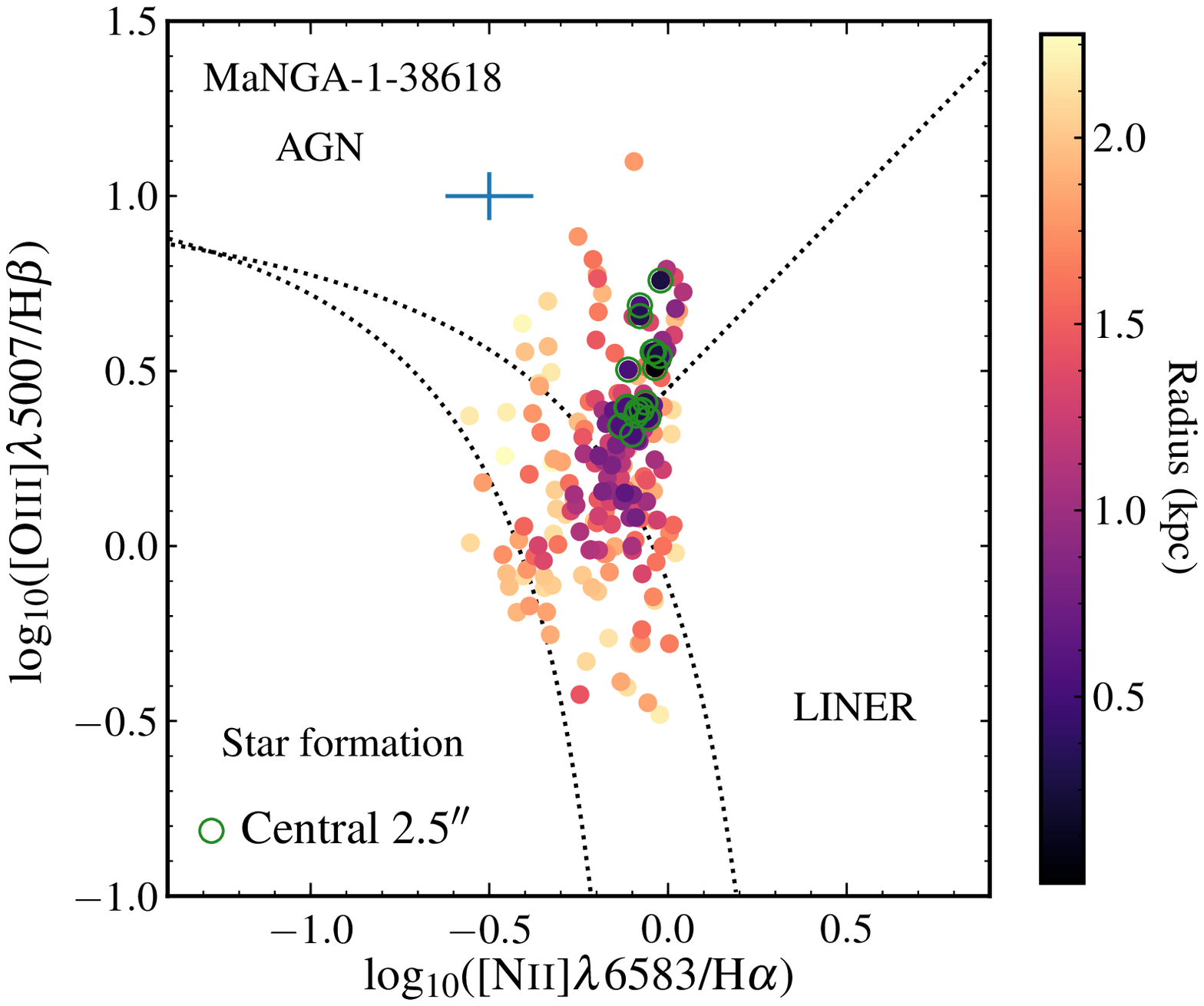}\\
\caption{Spaxel-by-spaxel BPT emission line ratio diagrams for two of the five galaxies in our sample with a kinematically-offset gas component. BPT diagrams for the remaining two galaxies are given in Appendix\,\ref{fig:appbptoff} The colour of each point corresponds to its distance from the galaxy centre, with the darkest points representing the central spaxels. Spaxels in the central 2.5\,arcsecs (the PSF of the reduced datacubes) are circled in green. No object in this sample exhibits line ratios consistent with solely being from ongoing star formation. Also plotted for diagnostic purposes are the \citet{2001ApJS..132...37K} and  \citet{2003MNRAS.346.1055K} classification lines, which are used to separate extreme starbursts and \ion{H}{ii} regions from AGN-like emission. The \citet{2007MNRAS.382.1415S} division between Seyfert-like and LINER-like (Low Ionisation Nuclear Emission Region) emission line ratios is also plotted.}
\label{emmaps}
\end{figure*}

\begin{figure*}
\centering
\includegraphics[width=0.45\textwidth]{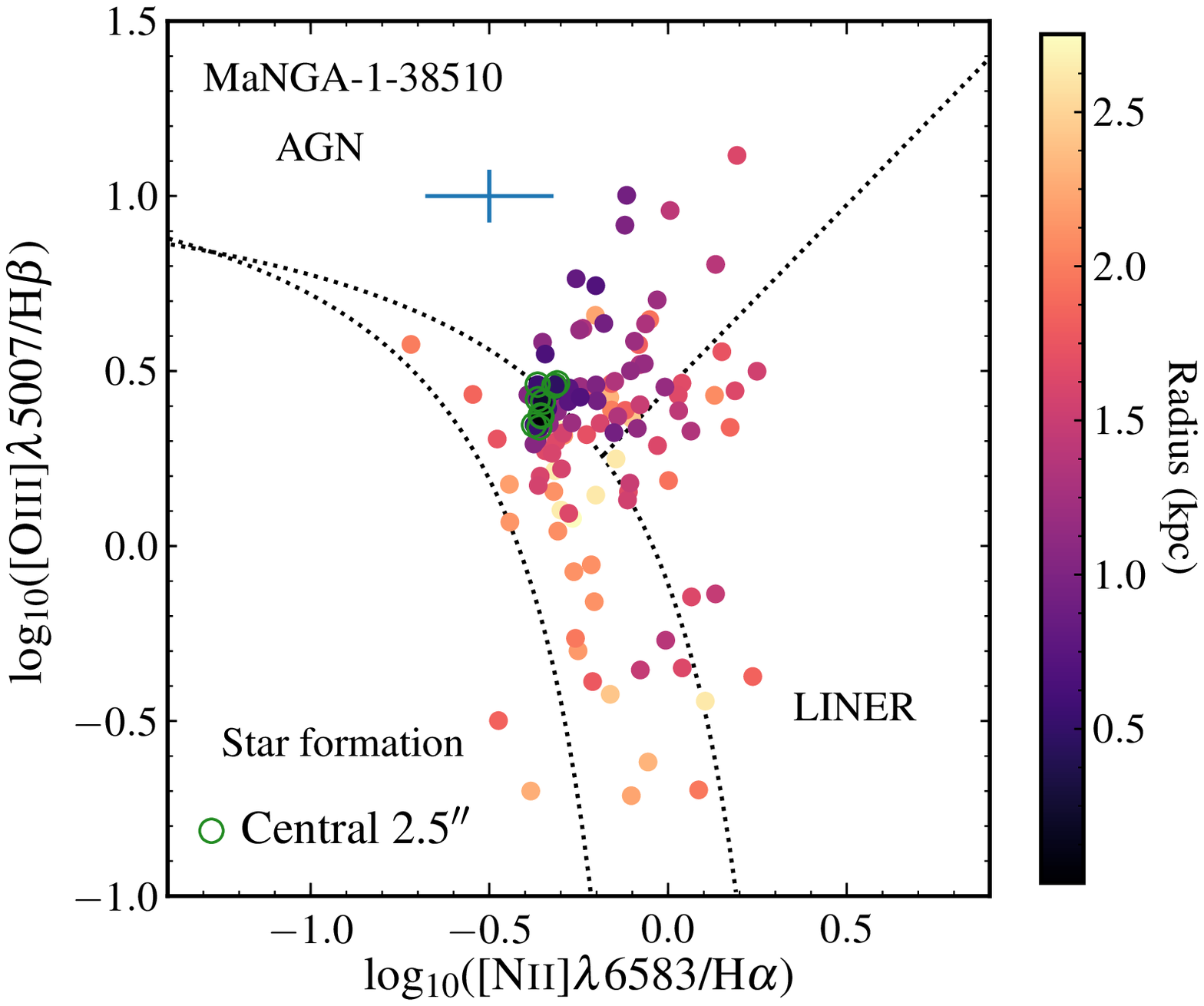}
\includegraphics[width=0.45\textwidth]{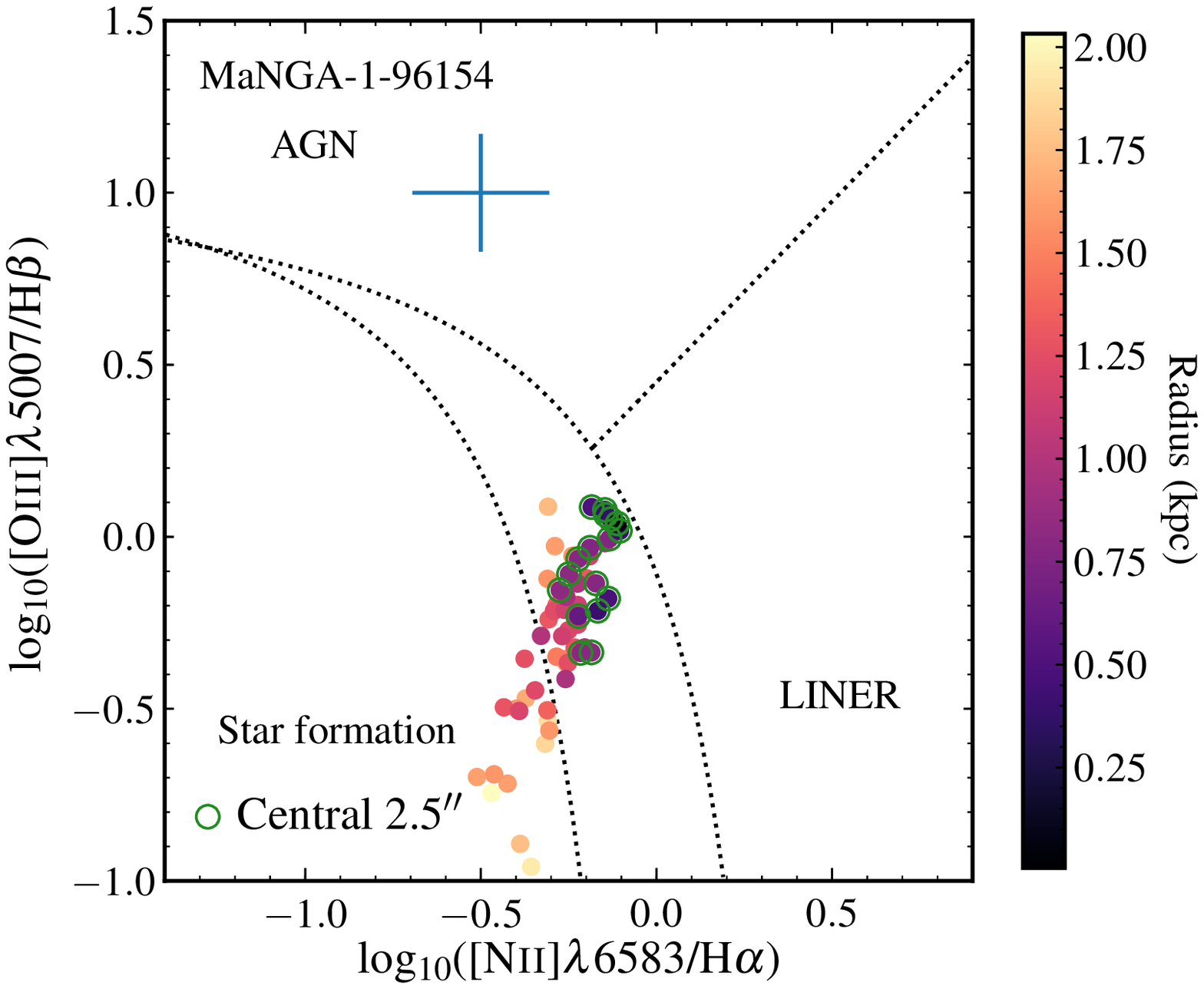}\\
\caption{Spaxel-by-spaxel BPT emission-line ratio diagrams for two galaxies in our sample with gas that is either co-rotating or offset by $\approx180\degr$ (counter rotating) from their stellar component. BPT diagrams for the remaining seven galaxies are given in Appendix\,\ref{fig:appbptco}. The colour of each point corresponds to its separation from the centre of the galaxy in arcseconds. Spaxels in the central 2.5\,arcsecs (the PSF of the reduced datacubes) are circled in green. The majority (8/9) of the galaxies with co- or counter-rotating gas have emission line ratios consistent with star formation or composite AGN/star formation. The remaining galaxy, MaNGA-1-230177, has central spaxels that exhibit emission line ratios consistent with AGN activity.} 
\label{emnooff}
\end{figure*}

We construct BPT diagrams for all objects in our quenched sample which exhibit an ionised gas component, to separate those with nuclear starbursts from those exhibiting AGN activity. The line ratios are calculated for each spaxel in the datacube with a signal-to-noise $>10$ in the stellar continuum, and a signal-to-noise $>5$ in their H$\alpha$ emission line. These emission line ratio diagrams are presented in Fig.~\ref{emmaps} (for those objects with kinematically offset rotation), and Fig.~\ref{emnooff} for those objects with co-rotating gas. Also plotted for diagnostic purposes are the \citet{2001ApJS..132...37K} and  \citet{2003MNRAS.346.1055K} classification lines classification lines, which are used to separate extreme starbursts and \ion{H}{ii} regions from AGN-like emission. The \citet{2007MNRAS.382.1415S} division between Seyfert-like and LINER-like (Low Ionisation Nuclear Emission Region) emission line ratios is also plotted, to ensure we identify genuine AGN emission in these galaxies, rather than heating by hot, old stars, which can masquerade as LINER emission \citep[e.g.][]{2011MNRAS.413.1687C,2016MNRAS.461.3111B}. Each spaxel is coloured by its distance from the galaxy centre, such that central spaxels are represented by darker points. 

Fig.\,\ref{emmaps} and Appendix\,\ref{fig:appbptoff} shows that the galaxies with kinematically offset gas exhibit AGN-like emission in their central regions, with no spaxels in any of the five galaxies consistent with current star formation. In contrast, eight of the nine galaxies plotted in Fig.\,\ref{emnooff} and Appendix\,\ref{fig:appbptco} with co-rotating or counter-rotating ionised gas have spaxel-by-spaxel BPT diagrams dominated by star formation or composite (AGN+star-formation) emission line ratios. Only one object with co-rotating stars and gas, MaNGA-1-230177, has clear AGN-like emission at its centre. We note that several galaxies with AGN-like emission ratios at their centres have spaxels consistent with LINER-like emission in their outer regions. This wide-spread LINER activity is consistent with either heating from the central AGN, or from old stellar populations in these outer regions. This wide-spread  LINER-like emission is not seen in those galaxies with central star formation or composite AGN/star formation emission line ratios.

\subsubsection{WHAN-classified AGN}

We also construct spaxel-by-spaxel BPT diagrams for the 22 galaxies classed as AGN using the WHAN classification system, but we do not include these plots. 12 objects in this sample exhibit central spaxels in the AGN/LINER/star formation composite region of the BPT diagram, five of which do not exhibit any purely star forming spaxels, while seven host star formation in their outer regions.  As such, these seven objects are not truly quenched galaxies, and have either blue colours $(u-r) < 1.65$, else are dust reddened edge-on spirals with axial ratios $b/a \leq 0.4$. Ten galaxies have line ratios at their centres consistent with active star formation. As we cannot classify any of these objects as clear AGN host candidates, we do not analyse them further in this work. 

\subsection{Emission-line maps}

\citet{2016Natur.533..504C} identified a number of galaxies in the MaNGA survey with AGN-powered maintenance-mode feedback, in which further star formation is prevented by the heating or removal of gas after the initial quenching episode. Such galaxies exhibit bi-symmetric emission line features, traced by the equivalent widths of their H$\alpha$ line. These  emission features are co-aligned with their ionised-gas velocity fields, and are found in $\sim10\%$ of MaNGA galaxies with stellar masses $\sim2\times10^{10}$\,M$_{\sun}$. 

We construct maps showing the distribution of the H$\alpha$ emission within the galaxy to search for such features in the low-mass galaxies with AGN-like emission.  The equivalent widths for the H$\alpha$ line are provided as part of the DAP output, and these maps are shown in Fig.\,\ref{ewmaps}. As can be seem in Fig.\,\ref{ewmaps}, four of the galaxies exhibit clear bi-symmetric emission features, though these features are not necessarily aligned with the velocity fields of their ionised gas. We also construct these maps for the galaxies with line ratios consistent with star formation, and these bi-symmetric features are not present in that sample. 

\begin{figure*}
\centering
\includegraphics[width=\textwidth]{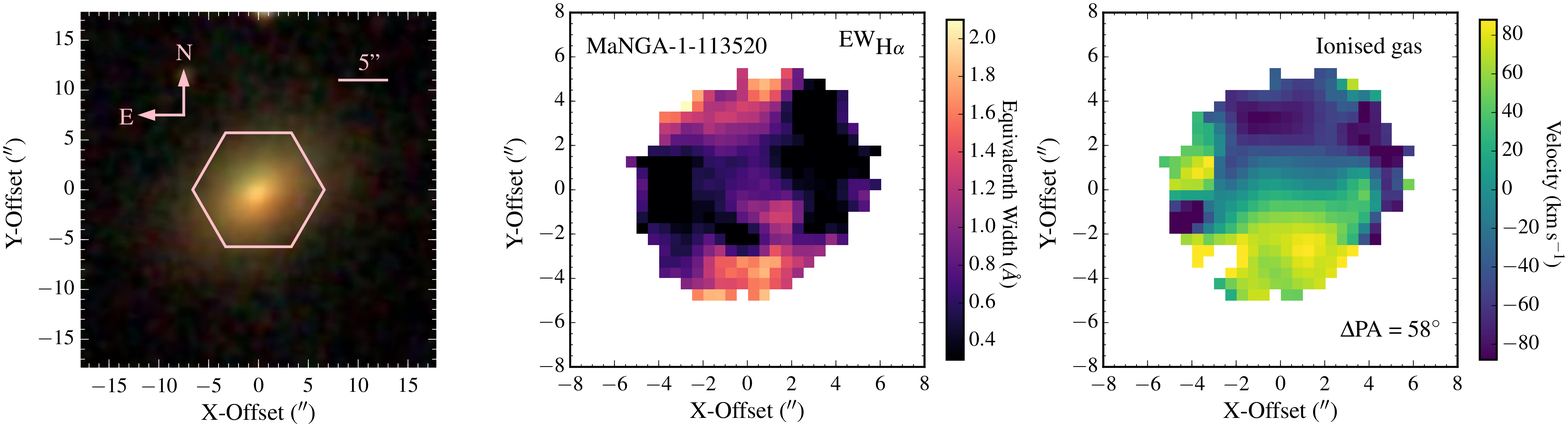}\\
\includegraphics[width=\textwidth]{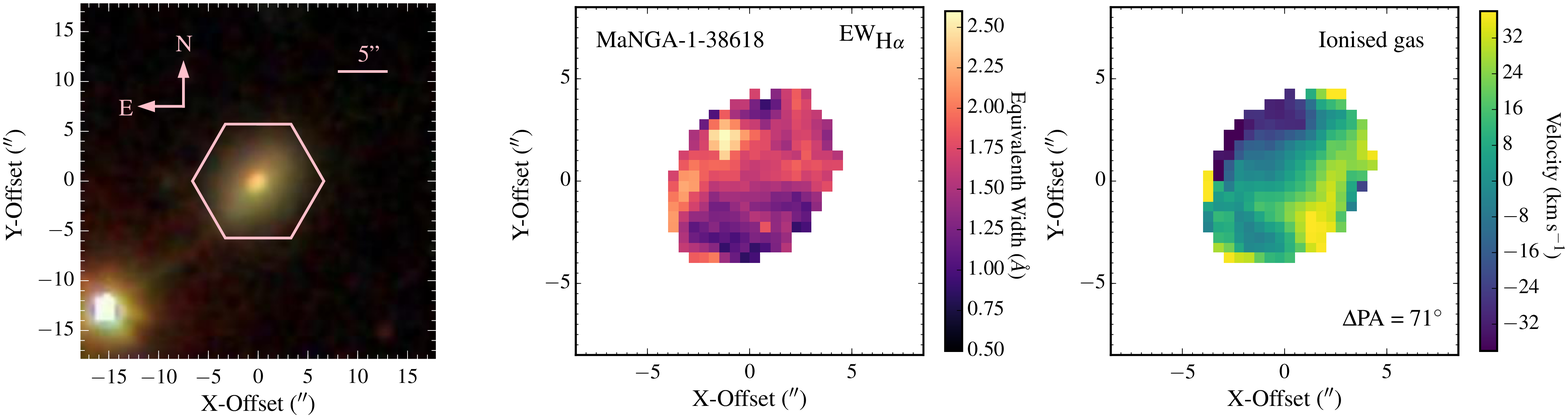}\\
\includegraphics[width=\textwidth]{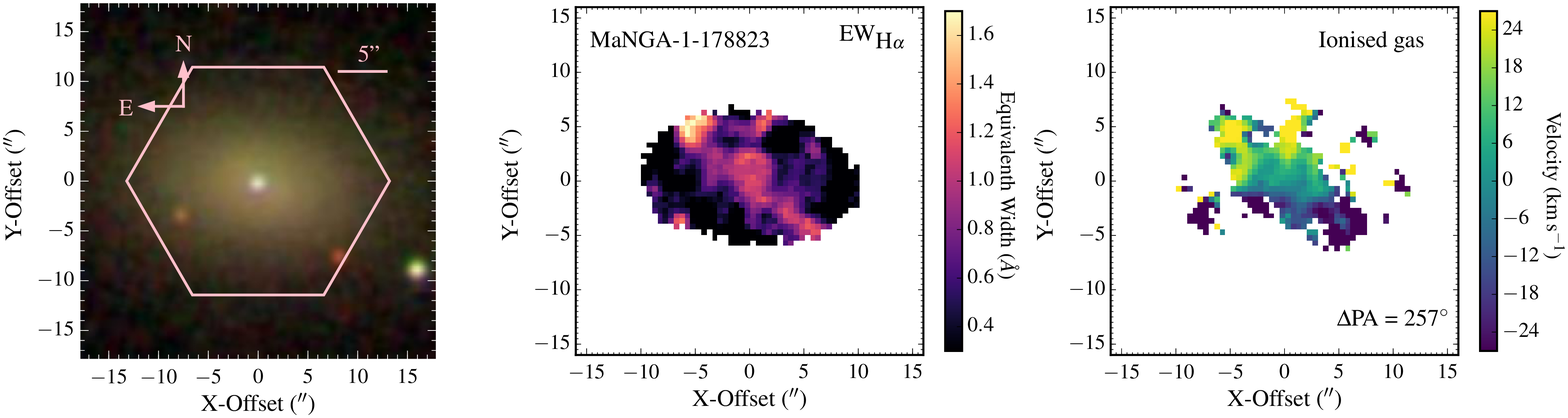}\\
\caption{H$\alpha$ equivalent maps for the six dwarfs in which AGN-like emission dominates at their centres (central panels). Their ionised gas velocity fields are shown for comparison (right-hand panels). The left-hand panels for each galaxy show an SDSS colour image, with the MaNGA IFU field-of-view overlaid as a hexagon.}
\label{ewmaps}
\end{figure*}

\begin{figure*}
\centering
\includegraphics[width=\textwidth]{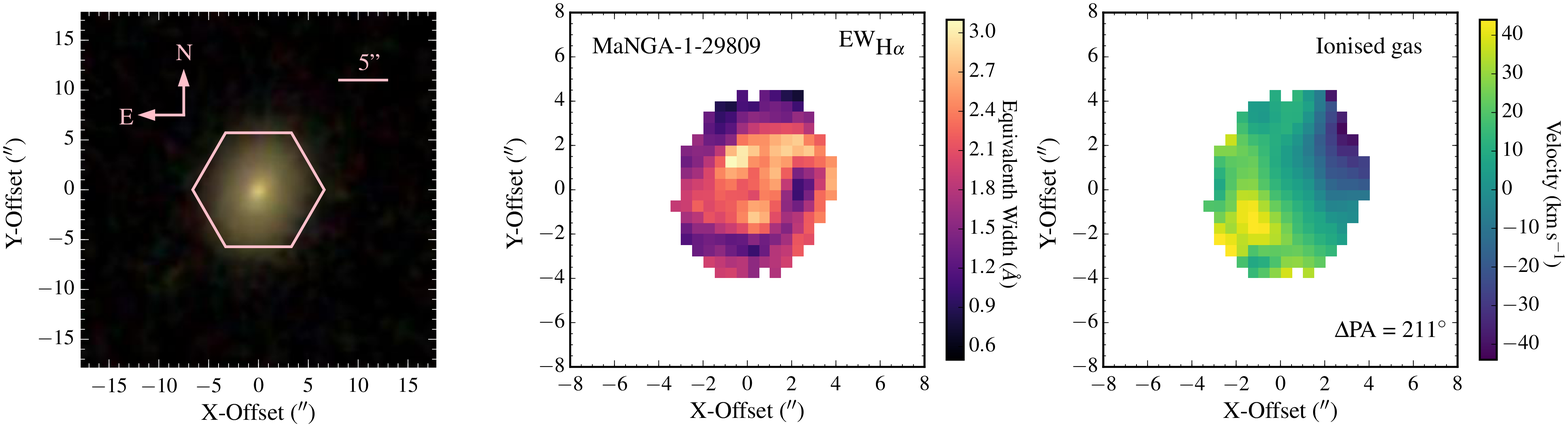}\\
\includegraphics[width=\textwidth]{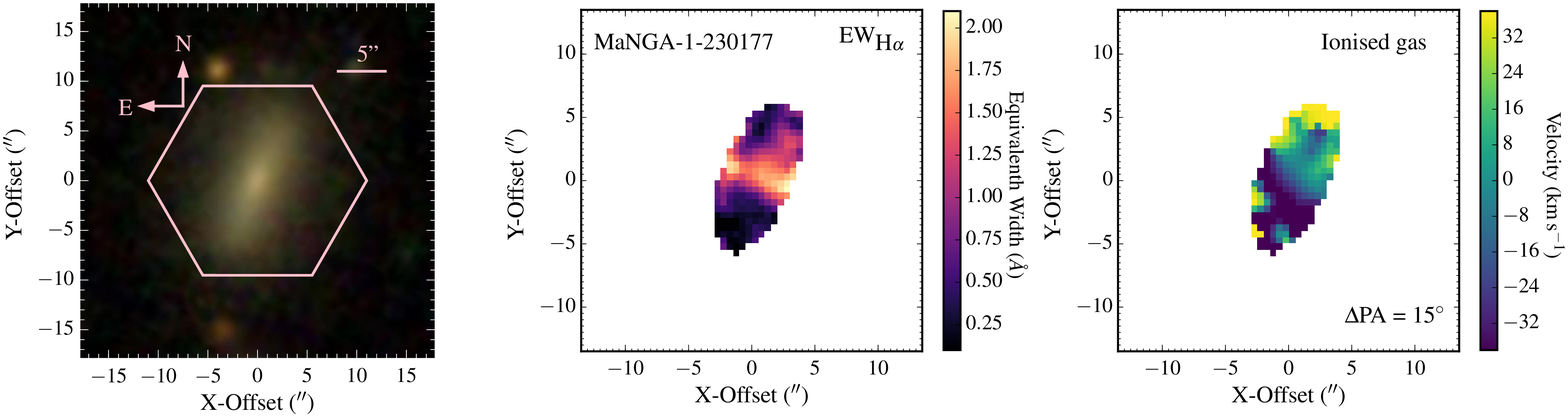}\\
\includegraphics[width=\textwidth]{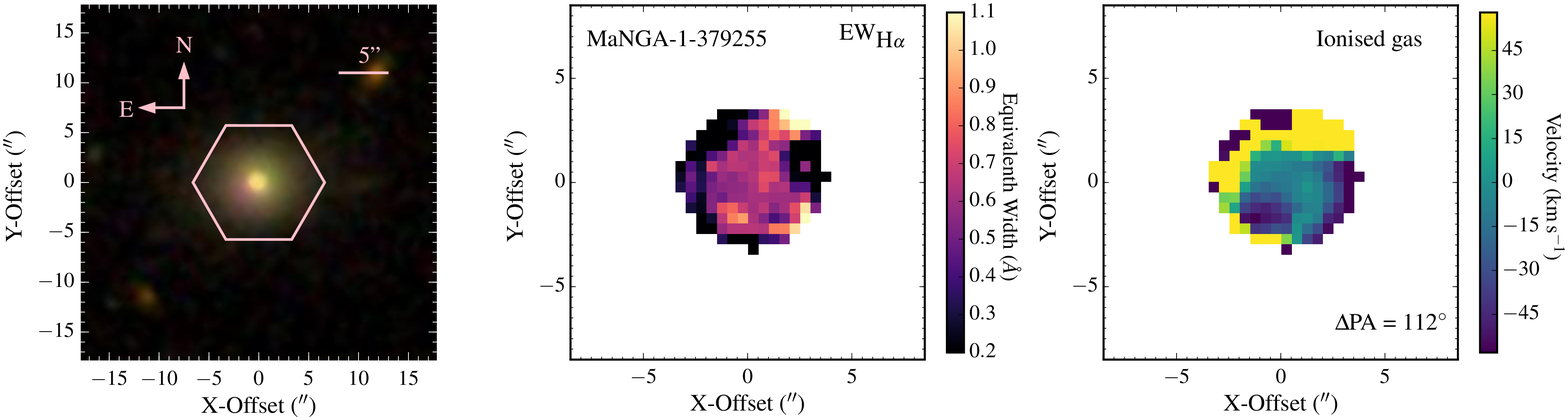}\\
\contcaption{H$\alpha$ equivalent maps for the six dwarfs in which AGN-like emission dominates at their centres.} 
\end{figure*}

\subsection{Mid-IR colours}
\label{sec:midir}

\begin{figure}
\includegraphics[width=\columnwidth]{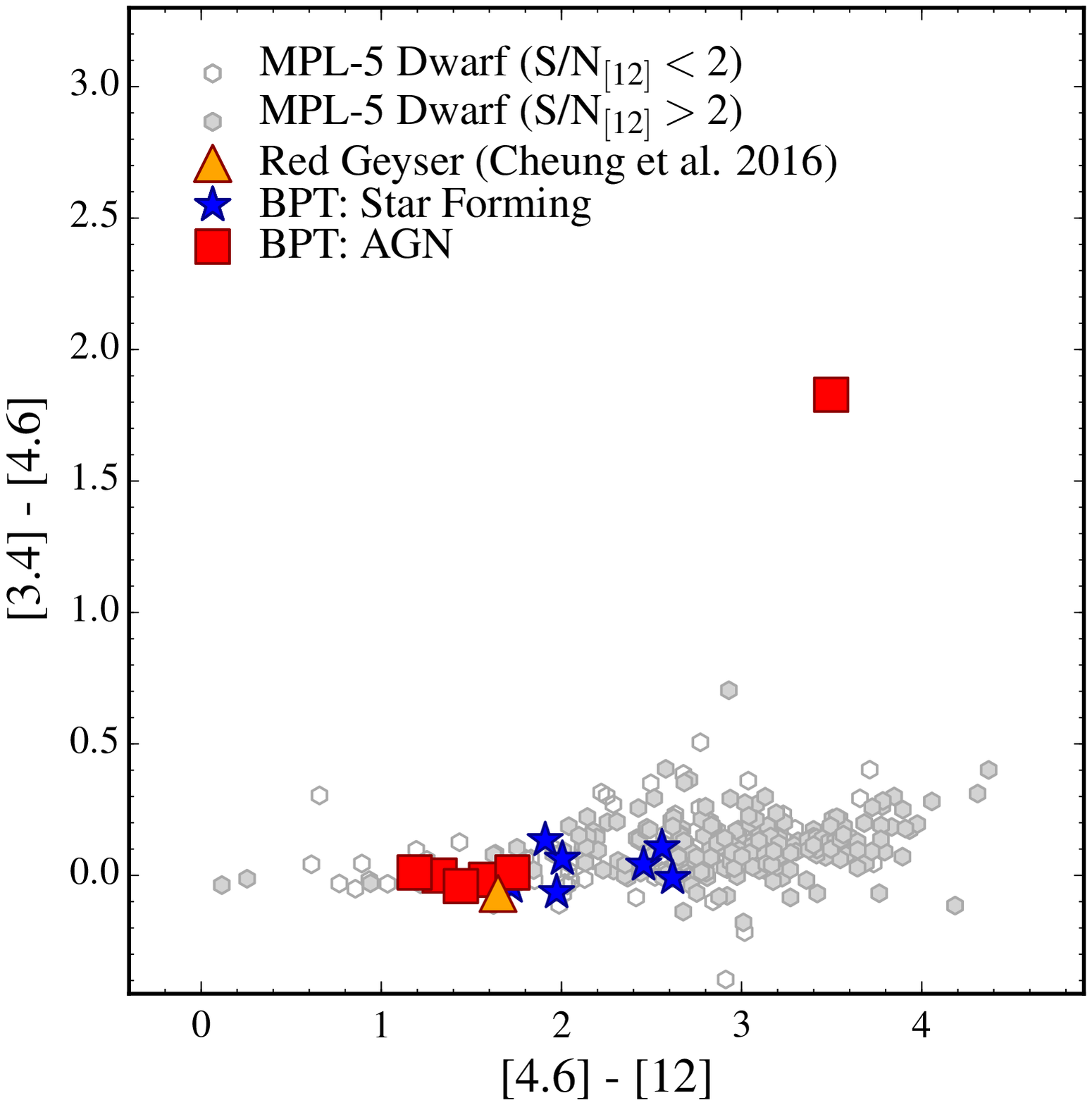}
\caption{\textit{WISE} colour-colour diagram for low-mass galaxies in MaNGA. The BPT-classified AGN hosts are shown as red squares. In general, the AGN-hosts exhibit bluer $[4.6]-[12]$ colours than those that host composite or star forming regions at their centres, consistent with little-to-no ongoing star formation. However, one low-mass galaxy, MaNGA-MaNGA 1-29809, has an extremely red$[3.4]-[4.6]$ colour, consistent with those of Seyfert-like hosts previously examined in the literature \citep[e.g.][]{2014MNRAS.438.1149G}.}
\label{fig:mir}
\end{figure}

Mid-IR photometry is a useful diagnostic tool for the identification of AGN in the absence of radio or X-ray data, particularly for objects hosting Seyfert-like emission, or obscured AGN.  
For example, radio-loud AGN are expected to radiate brightly in the mid-IR if the dust torus of the AGN strongly obscures the central black hole. 
Low-excitation radio galaxies hosting AGN are harder to identify in the mid-IR however, and exhibit a range of mid-IR colour \citep[e.g.][]{2014MNRAS.438.1149G}. 

Mid-IR colours can also be used to separate out quenched stellar populations from those with recent star formation. For example, red-sequence, optically quenched galaxies that have undergone star formation in the past 2\,Gyr exhibit excess mid-IR emission \citep[e.g.][]{2013ApJ...767...90K}. 
This is due to emission from the dusty circumstellar envelopes of asymptotic giant branch (AGB) stars, and polycyclic aromatic hydrocarbon (PAH) release from carbon stars. 
The contribution of these processes to the mid-IR emission of a galaxy decreases as the mean stellar age of the galaxy increases. 

We confirm the current star formation state of the galaxies using mid-IR photometry taken from the \textit{WISE}  telescope \citep{2010AJ....140.1868W} All Sky Survey. We plot the $[3.4]-[4.6]$ versus $[4.6]-[12]$ colours of the entire low-mass galaxy sample in Fig.\,\ref{fig:mir}. Those objects with $S/N <2$ in the \textit{WISE} are shown as unfilled circles. We also plot the positions of the objects classed as AGN or star forming via the line-strength diagnostic techniques presented in Section\,\ref{sec:bpt}. One object in our AGN sample, MaNGA 1-38618, is excluded from this plot, as its photometry is flagged as unreliable by the \textit{WISE} photometry pipeline, likely due to its close proximity to a foreground star (\textit{WISE} has an angular resolution of $\sim$\,arcsec). 

One galaxy, MaNGA 1-29809, lies in the region of the \textit{WISE} $[3.4]-[4.6]$ versus $[4.6]-[12]$ colour-colour diagram typically occupied by Seyfert galaxies and obscured AGN \cite[see figure 12 of ][]{2010AJ....140.1868W}, with $[4.6]-[12] = 3.50$, and $[3.4]-[4.6] = 1.83$. 
The red $[3.4]-[4.6]$ colour indicates a heavily obscured Type-2 AGN \citep[][]{2011ApJ...735..112J}.
The remaining BPT-classified AGN have $ 1.19 < [4.6]-[12] < 1.73$, consistent with them hosting little or no star formation, and they likely host primarily quenched stellar populations. 
Never-the-less, they are not as blue as many elliptical galaxies, which typically have $0 < [4.6]-[12]< 1.5$. 
In contrast, those galaxies which exhibit composite or star-formation like emission line ratios are typically redder, with $1.72 < [4.6]-[12] < 2.62$, and lie in the region of the colour-colour diagram occupied by star forming galaxies. 
$[4.6]-[12]$ colours indicate the amount of dust heating in a galaxy due to star formation, with bluer colours due to little-to-no star formation. 
Though we do not plot them in Fig.\,\ref{fig:mir} for simplicity, those galaxies identified as AGN using the WHAN classification system have mid-IR $[4.6]-[12]$ colours consistent with active star formation. 

We also checked the Faint Images of the Radio Sky at Twenty Centimeters \citep[FIRST][]{1995ApJ...450..559B}) radio survey for detections of our AGN-host galaxies, but no sources were found within 10\,arcsec of the AGN-host candidates. Stacked FIRST imaging for the AGN-host candidates also did not reveal a detection. A deeper radio continuum survey is therefore required to constrain the individual radio powers of the galaxies. No X-ray data is available for these objects.

\subsection{Star formation histories}

We check the spatially resolved star formation histories for the AGN hosts, to see if any have undergone recent star formation that has been quenched by the AGN. The mean stellar ages of the objects are provided by the MaNGA Pipe3D pipeline, which provides stellar population fits to all MaNGA galaxies using simple stellar population models. For more information on MaNGA Pipe3D, see \citet{2016RMxAA..52...21S,2016RMxAA..52..171S}.

In the lowest redshift object in our sample, MaNGA\,1-178823, we were able to resolve recent star formation within its central 0.25\,kpc. The galaxy has a mean stellar age $<1\pm0.8$\,Gyr with its central 0.25\,kpc, suggesting a recently quenched nuclear starburst. Within 1\,$R_{e}$, MaNGA\,1-178823 also exhibits a clear post-starburst spectrum with A-star like spectra with strong H$\delta$ and H$\beta$ absorption. Young central ages $<1.5$\,Gyr are also found for MaNGA\,1-29809 and MaNGA\,1-38618, again showing they have undergone recently quenched star formation episodes. Beyond their central regions, the stellar populations of these galaxies are typically old, with mean stellar ages $>3$\,Gyr. The remaining three AGN-host objects are dominated by old stellar populations with ages $>3$\,Gyr throughout their structures.

\subsection{AGN luminosities}

We estimate the AGN luminosities for our sample using their [O\textsc{iii}]$\lambda$5007 emission lines. The [O\textsc{iii}] fluxes $F_{\textrm{[O\textsc{iii}]}}$ within the central 0.5\,kpc radius of each dwarf are converted to luminosities using $L_{\textrm{[O\textsc{iii}]}} = 4 \pi D_{L}^{2}  F_{\textrm{[O\textsc{iii}]}}$, where $D_{L}$ is the luminosity distance of the galaxy. We do not correct the fluxes for dust reddening internal to the galaxies, and we assume the flux is dominated by the contribution from the AGN.

The dwarfs have $1.5\times10^{37} \textrm{erg\,s$^{-1}$} <L_{\textrm{[O\textsc{iii}]}} < 1.64\times10^{38}$\,erg\,s$^{-1}$, with median $L_{\textrm{[O\textsc{iii}]} }= 1.05\times10^{38}$\,erg\,s$^{-1}$. We convert this value of $L_{\textrm{[O\textsc{iii}]}}$ to a bolometric luminosity $L_{bol}$ using the bolometric correction  $L_{bol} \approx 3500L_\textsc{[O\textsc{iii}]}$ from \citet{2004ApJ...613..109H}.  This gives a typical AGN bolometric luminosity $L_{bol}=3.7\times10^{41}$\,erg\,s$^{-1}$ for the low-mass AGN host galaxies in our sample. However, the bolometric correction we have used may not be applicable to these low-mass systems, as its calibration was done using galaxies of much higher mass. \citet{2014AJ....148..136M} suggest a bolometric correction $L_{bol}/L_\textsc{[O\textsc{iii}]} = 1000$ for dwarf AGN systems, so the AGN bolometric luminosities may be as low as  $L_{bol}=1.06\times10^{41}$\,erg\,s$^{-1}$.

We also estimate a typical black hole mass for these system following \citet{2013ApJ...764..184M}, though without reliable measures of $\sigma_{\star}$ for the majority of our objects, we note this is an upper limit. We assume $\sigma_{\star} \approx 50$\,km\,s$^{-1}$ for the low mass galaxies in our sample, typical for dEs in their luminosity range \citep[e.g.][]{2016MNRAS.462.3955P}, and consistent with the recovered values of $\sigma_{\star}$ presented in Table\,\ref{tab:kinpas}. This gives a typical black hole mass log$_{10}$($M_{BH}/M_{\sun}) \approx 5$ for these low-mass galaxies. 

\section{Discussion}
\label{sec:discuss}

\subsection{Low-mass AGN-host galaxies}

We identify six low-mass, non-star forming galaxies that exhibit AGN-like emission line ratios in their spatially-resolved BPT diagrams (Figs. \ref{emmaps} and \ref{emnooff}). Of these, six have kinematically offset gas, such that the velocity field of their ionised gas component is offset by $>30\degr$ from the kinematic PA of their stellar component (Fig.\,\ref{fig:vfields}). The remaining galaxy with central AGN-like emission has an ionised gas component that is co-rotating with the stars. All six candidates have $M_{\star} \lesssim 5\times10^{9}$\,M$_{\sun}$, and $M_{r} > - 19$, placing them in the dwarf galaxy regime. 

Using the luminosities of their O[\textsc{iii}] emission lines, we find $L_{\textrm{O\textsc{iii}}}\approx1.05\times10^{38}$\,erg\,s$^{-1}$ for a typical low-mass galaxy in our sample ($L_{bol}=3.7\times10^{41}$\,erg\,s$^{-1}$). These luminosities are comparable to those identified for similar luminosity galaxies in the literature \citep[e.g.][]{2014AJ....148..136M}, suggesting AGN may play a role in low-mass galaxy evolution. Further observations are required to establish if evidence for maintenance mode feedback, such as bi-symmetric emission line features, are common in these low-mass AGN-host galaxies. 

Based on the kinematic offset of the gas in the majority of the dwarfs with nuclear AGN-like emission, and the large velocity offset from the stellar kinematics, we infer that this gas is either being accreted onto, or expelled from, the low-mass galaxies. As these low-mass galaxies also exhibit AGN-like emission line ratios, we suggest that maintenance-mode AGN feedback is preventing the gas in these galaxies from cooling and triggering new bursts of star formation. These faint galaxies are therefore likely low-mass analogues of the ``red geysers'' presented in \citep{2016Natur.533..504C}, which maintain quiescence through low-level AGN heating which prevents accreted gas from cooling and forming new stars.  Two galaxies in particular, MaNGA 1-113520 and MaNGA 1-230177, do not contain any spaxels in their MaNGA datacube with emission line ratios consistent with photoionisation from star formation. A number of these low-mass AGN host galaxies also exhibit mean stellar ages $<1.5$\,Gyr at their centres, showing they have undergone a recent quenching episode, likely the result of this AGN feedback.

We discuss the properties of individual galaxies in the sample with evidence for AGN feedback below. 

\subsubsection{MaNGA 1-38618} 

MaNGA 1-38618 has a stellar mass $2.8\times10^{9}$\,M$_{\sun}$, and $M_{r} = -18.65$. It exhibits weak H$\alpha$ emission throughout its structure, with no clear symmetric features and a kinematic position angle offset between its gas and stars of $\sim90\degr$. It has emission line ratios in its central regions consistent with AGN/LINER activity, while at larger radii the emission is composite between star formation and LINER-like. It also exhibits a young central stellar population, with mean stellar ages $<1.5$\,Gyr, suggesting it has undergone a recently quenched star formation episode.

\subsubsection{MaNGA 1-379255} 

MaNGA 1-379255 has a stellar mass of $3.03\times10^{9}$\,M$_{\sun}$. The galaxy has a kinematic offset of $77\degr$ between its gas and stars, and contains central spaxels which lie on the region of the BPT diagram occupied by AGN-like galaxies. Its emission features are not bi-symmetric. However, it has few spaxels above our signal-to-noise threshold.

\subsubsection{MaNGA 1-230177} 

MaNGA 1-230177 is unusual in our sample of six dwarf-like AGN hosts, in that its gaseous and stellar components are co-rotating, with a kinematic offset of $15\degr$ between the two. It has a stellar mass $3.98\times10^{9}$\,M$_{\sun}$, and it has bar-like in shape with no spiral arms in \textit{gri} colour imaging. The galaxy also exhibits very clear bi-symmetric H$\alpha$ emission aligned with its minor axis. 

\subsubsection{MaNGA 1-178823}

MaNGA 1-178823 is the lowest redshift object in our AGN sample, with $z = 0.013$ (48\,Mpc). It has a stellar mass $4.39\times10^{9}$\,M$_{\sun}$, and $M_{r} = -18.92$. MaNGA 1-178823 has a bright, point-like nucleus clearly visible in its SDSS colour imaging, consistent with emission from an AGN, or the nuclear star clusters seen in bright dwarf galaxies in galaxy clusters like Fornax \citep[e.g.][]{2015ApJ...813L..15M}. It has very obvious bi-symmetric emission, and  like the majority of galaxies in our AGN sample, it exhibits a large kinematic offset between its gas and stars, such that the rotation of its ionised gas emission as traced by the H$\alpha$ emission line is offset by $61\degr$ from its stellar component. It also exhibits weak $D4000$ absorption at its centre (i.e. younger mean stellar ages) compared to outer regions, with mean stellar ages in its central 0.25\,kpc $<1$\,Gyr. MaNGA 1-178823 is located in the NGC 6962 group, at a projected distance of just $16$\,kpc and $122$\,km\,s$^{-1}$ from the S0 galaxy NGC 6959, and the two galaxies are therefore likely tidally interacting. The BPT diagram of the galaxy is primarily dominated by LINER-like emission to large radii, suggesting the presence of old, low-mass stars, though the nuclear region exhibits Seyfert-like emission. Based on line-strength ratios, it does not exhibit any star formation throughout its structure. 

\subsubsection{MaNGA 1-113520} 

MaNGA 1-113520 does not exhibit any emission lines in the star forming part of the BPT diagram, with the most central spaxels exhibiting AGN emission line ratios. The galaxy exhibits a counter-rotating core in its stellar component, and its gas is kinematically offset from the stellar components by $\sim68\degr$. It furthermore exhibits bi-symmetric H$\alpha$ emission, with this emission spatially aligned with the kinematically-offset gas component \ref{ewmaps}. The equivalent widths of the H$\alpha$ features increase with radius from the galaxy nucleus, and resembles several cartoon models of AGN feedback.  It is not detected in MaNGA Green Bank \ion{H}{i} single dish follow-up observations (Masters et al., in prep.). The galaxy is not detected at either radio or X-ray wavelengths, and has mid-IR colours  $[3.4]-[4.6] = 1.6$ and $[4.6]-[12] = 0.0$, consistent with the galaxy hosting little star formation in the past 2\,Gyr. 

\subsubsection{MaNGA 1-29809}

Of the six galaxies with AGN-like emission, one galaxy, MaNGA 1-29809 shows evidence across multiple wavelengths for hosting an active nucleus. The strongest H$\alpha$ emission is centrally located, with a hint of a ring-like structure at larger radii. It has \textit{WISE} mid-IR colours $[3.4]-[4.6] = 1.8$ and $[4.6]-[12] = 3.5$, placing it in the region of the colour-colour diagram occupied by Seyferts, ULIRGS, and LINERS. This object was also identified as a dwarf AGN by \citet{2015MNRAS.454.3722S} based on this $[3.4]-[4.6]$ mid-IR colour. It is offset from a point-like source in the radio-continuum FIRST survey  by $\sim30''$, with a flux $3.19\pm0.097$~mJy, however this is likely to be a background source. No obvious galaxy or star is detected at optical wavelengths is detected at this location. No X-ray observations are available for this galaxy. 

With a stellar mass $3.6\times10^{9}$\,M$_{\sun}$, it is one of the lowest mass galaxies with clear signatures of hosting an AGN identified to date.  MaNGA 1-29809 is fairly well isolated, located at a distance $D = 928$\,kpc and $\Delta v = 799$\,km\,s$^{-1}$ from the S0 galaxy IC 1517 ($M_{K} = -24.6$).  The central velocity dispersion of MaNGA 1-29809 is well below the BOSS spectrograph's instrumental resolution, and thus we cannot easily calculate the mass of its supermassive black hole using the central velocity dispersion. However, we can place limits on this mass, using $\sigma_{\rm{cen}} = 50$\,km\,s$^{-1}$ as an upper limit on the galaxy's central velocity dispersion. With a central velocity dispersion $\sigma_{\rm{cen}} = 50$\,km\,s$^{-1}$, this would imply a central black hole mass $<10^{6}$\,M$_{\sun}$, assuming MaNGA 1-29809 lies on the $M - \sigma$ relation \citep[e.g.][]{2000ApJ...539L...9F}. However, the resolution of the data are not sufficient to examine the broad line regions of the AGN emission. The galaxy has a mean stellar age $<1.5$\,Gry with its central 0.25\,kpc, suggesting it has been recently quenched after a star formation episode. 

\subsection{Low-mass galaxies hosting star formation}

The remaining eight galaxies in our sample do not exhibit emission line features consistent with AGN activity. Instead, their spatially resolved BPT diagrams are dominated by line ratios suggesting they host ongoing star formation. This star formation is typically located in the outer regions of the galaxies, and most (7 out of 8) have central spaxels that reside in the star-formation/AGN/LINER composite region of the BPT diagram, i.e the region between the \citet{2001ApJS..132...37K} and  \citet{2003MNRAS.346.1055K} diagnostic lines. We cannot rule out these galaxies as AGN hosts, as both AGN activity and star formation is contributing to the heating of their ionised gas component. 

\subsection{AGN host galaxy fraction}

10\,per\,cent of the galaxies in our quenched dwarf galaxy sample appear to host genuine AGN, based on BPT diagnostics. This is comparable to the X-ray identified AGN fraction identified in previous work for low mass galaxies with M${\star} > 5\times10^{9}$\,M$_{\sun}$ \citep{2012ApJ...746...90A,2016A&A...588A..78B}. We also compare the fraction of low-mass galaxies in our sample hosting AGN, to that of the general galaxy population over the same mass range. Restricting our sample to a stellar mass range $10^{9}$\,M$_{\sun}$~$<$~M$_{\star} < 5\times10^{9}$\,M$_{\sun}$, and including the dwarfs with composite AGN/SF emission at their centres, we have a total of 13 possible dwarf AGN host galaxies.  248 low-mass MaNGA galaxies fall in the same mass range. We can therefore place an upper limit of 5\,per~cent of all MaNGA galaxies at this mass range hosting AGN, while the ratio of AGN host galaxies to star forming galaxies is $\sim7$ per\,cent. 

This compares to 10\,per~cent of AGN hosts across the mass range identified in \citet{2016A&A...588A..78B} calculated using estimates of the AGN duty cycle, and $\sim5$\,per~cent for galaxies with stellar masses $5\times10^{9}\textrm{ M$_{\sun}$ } <\textrm{M$_{\star}$} <  10^{10}$\,M$_{\sun}$ \citep{2012ApJ...746...90A}. For more luminous galaxies, the fraction of AGN hosts increases to 20-30 per-cent at low redshift \citep[$z=0.25$][]{2017MNRAS.469.3232G}. Thus our AGN fraction appears consistent with that in the literature for other low-mass galaxies.

\subsection{Alternative heating mechanisms}

AGN-like emission line ratios do not necessarily mean the gas in that galaxy has been ionised by a central supermassive black hole. Several other heating mechanisms exist within galaxies that can result in such ratios, including heating by hot, old, evolved stars (e.g. post-AGB stars). \citet{2016MNRAS.461.3111B} demonstrated that extended LINER-like emission found in many MaNGA galaxies is likely the result of ionisation by post-AGB stars. Post-AGB stars are the main source of ionising photons upon the cessation of star formation \citep{2008MNRAS.391L..29S,2011MNRAS.413.1687C}, though it is unclear if post-AGB stars are able to power the observed equivalent widths of LINER-like host galaxies due to the difficulty in modelling post-AGB stars.  \citet{2016MNRAS.461.3111B} refer to galaxies with LINER-like line-strength ratios resulting from heating by old stars as LIERs, as they masquerade as LINERS on the BPT diagram. However, the AGN-candidates in our sample have Seyfert-like emission at their centres, unlike that seen in the the LIER galaxies presented in \citet{2016MNRAS.461.3111B}. 

The mid-IR colours of all but one of the BPT-classified AGN are blue ($[4.6]-[12] < 1.7$), showing a lack of recent star formation throughout their structures. These line ratios might therefore be the result of feedback from hot, evolved stars.  We have assumed in this work that gas is either being driven in or out of the galaxies, due to the different kinematic properties of the stellar and ionised gas components. This kinematic offset shows the gas is not in dynamical equilibrium with the rest of the galaxy. Furthermore, several of the galaxies exhibit elevated H$\alpha$ velocity dispersions along their bi-symmetric emission line features, indicating gas outflows. However, higher spatial and instrumental resolution IFU data is needed to fully examine the nature of this emission. 

\section{Conclusions}
\label{sec:conclude}

We present evidence for AGN feedback in low-mass galaxies. Our sample, which is drawn from the first 2 years of the SDSS-IV MaNGA survey, provides spatially resolved conformation that low-mass galaxies which exhibit evidence for gas inflow/outflow are prevented from forming stars via AGN feedback. These are likely low-mass analogues of the ``red geysers'' presented in \citep{2016Natur.533..504C}. This result suggests that AGN feedback may play an important feedback role in low mass galaxies with $M_{\star} \sim 10^{9}$\,M$_{\sun}$. This feedback is likely most important in the galaxy group environment  in order to stop dEs with $M_{\star} \sim 10^{9}$\,M$_{\sun}$ undergoing further star formation after an initial quenching episode through e.g. tidal interactions. We summarise our conclusions below. 

\begin{itemize}
\item A key result of these study is that maintenance-mode AGN feedback may be required for some bright dE galaxies to maintain quiescence in galaxy groups and the outskirts of clusters. Environmental quenching is likely still the dominant quenching mechanism for dE galaxies, though AGN feedback may help some dEs maintain quiescence after the initial quenching episode. 
\item Five low-mass galaxies in our sample exhibit ionised gas components as traced through the H$\alpha$ emission line that are kinematically offset by $\gtrsim30\degr$ from their stellar velocity field. This means the gas in these galaxies is not in dynamical equilibrium, and it is either a recently accreted component, else an outflow. 
\item All five of galaxies with kinematically offset gas exhibit AGN-like emission line ratios at their centres, along with one galaxy that has an ionised gas  component that is co-rotating with its stars. 
\item One low-mass galaxy, MaNGA\,1-29809, exhibits evidence at optical and mid-IR wavelengths for hosting an AGN with Seyfert-like emission line ratios. With a stellar mass $3.6\times10^{9}$\,M$_{\sun}$, this is one of the lowest-mass AGN host galaxies identified to date. This makes it an ideal candidate for multi-wavelength follow-up observations, including both high-resolution imaging and more detailed IFU spectroscopy. 
\end{itemize}

While it is usually assumed that AGN feedback is unimportant in low-mass galaxy evolution, we have demonstrated here that $\sim10$\,per\,cent of galaxies in MaNGA with stellar masses $\lesssim5\times10^{9}$\,M$_{\sun}$ are AGN host candidates. Detailed observational followup is therefore needed to fully understand the role of feedback in these galaxies. 

\section*{Acknowledgements}

Funding for the Sloan Digital Sky Survey IV has been provided by the
Alfred P. Sloan Foundation, the U.S. Department of Energy Office of
Science, and the Participating Institutions. SDSS acknowledges
support and resources from the Center for High-Performance Computing at
the University of Utah. The SDSS web site is www.sdss.org.

SDSS is managed by the Astrophysical Research Consortium for the Participating Institutions of the SDSS Collaboration including the Brazilian Participation Group, the Carnegie Institution for Science, Carnegie Mellon University, the Chilean Participation Group, the French Participation Group, Harvard-Smithsonian Center for Astrophysics, Instituto de Astrofísica de Canarias, The Johns Hopkins University, Kavli Institute for the Physics and Mathematics of the Universe (IPMU) / University of Tokyo, Lawrence Berkeley National Laboratory, Leibniz Institut für Astrophysik Potsdam (AIP), Max-Planck-Institut für Astronomie (MPIA Heidelberg), Max-Planck-Institut für Astrophysik (MPA Garching), Max-Planck-Institut für Extraterrestrische Physik (MPE), National Astronomical Observatories of China, New Mexico State University, New York University, University of Notre Dame, Observatório Nacional / MCTI, The Ohio State University, Pennsylvania State University, Shanghai Astronomical Observatory, United Kingdom Participation Group, Universidad Nacional Autónoma de México, University of Arizona, University of Colorado Boulder, University of Oxford, University of Portsmouth, University of Utah, University of Virginia, University of Washington, University of Wisconsin, Vanderbilt University, and Yale University.

This research made use of Marvin \citep[][]{brian_cherinka_2017_292632}, a core Python package and web framework for MaNGA data, developed by Brian Cherinka, José Sánchez-Gallego, and Brett Andrews. (MaNGA Collaboration, 2017). This research made use of Astropy, a community-developed core Python package for Astronomy (Astropy Collaboration, 2013). This project makes use of the MaNGA-Pipe3D dataproducts. This project make use of the MaNGA-Pipe3D dataproducts. We thank the IA-UNAM MaNGA team for creating it, and the ConaCyt-180125 project for supporting them. We thank the referee for taking time to review our manuscript, and their suggestions led to the improvement of the paper.



\bibliographystyle{mnras}
\bibliography{spenny}



\appendix

\section{Stellar and ionised gas velocity maps}
\label{appendix:vmaps}

Stellar and ionised-gas velocity maps are provided in Fig.\,\ref{fig:appvfields} for the low-mass MaNGA galaxies examined in this work. The ionised gas velocity field is traced using the H$\alpha$ emission line. 

\begin{figure*}
\centering
\includegraphics[width=0.8\textwidth]{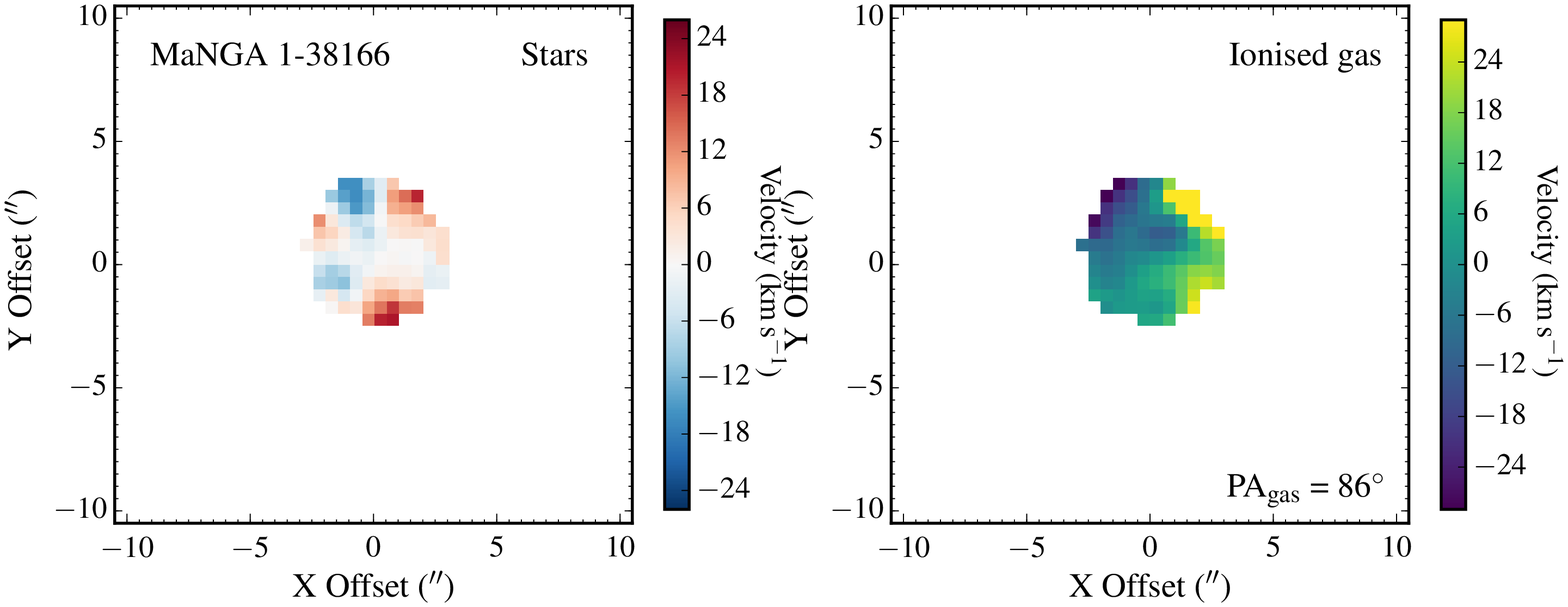}\\
\includegraphics[width=0.8\textwidth]{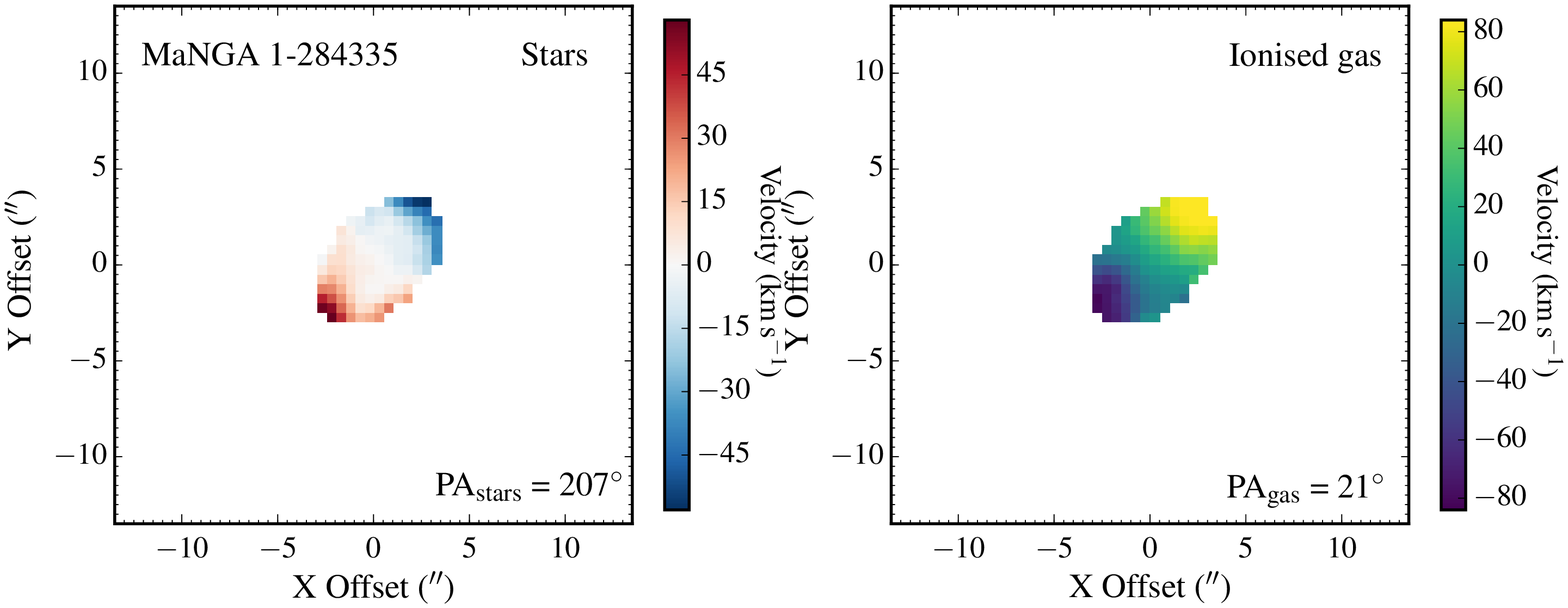}\\
\includegraphics[width=0.8\textwidth]{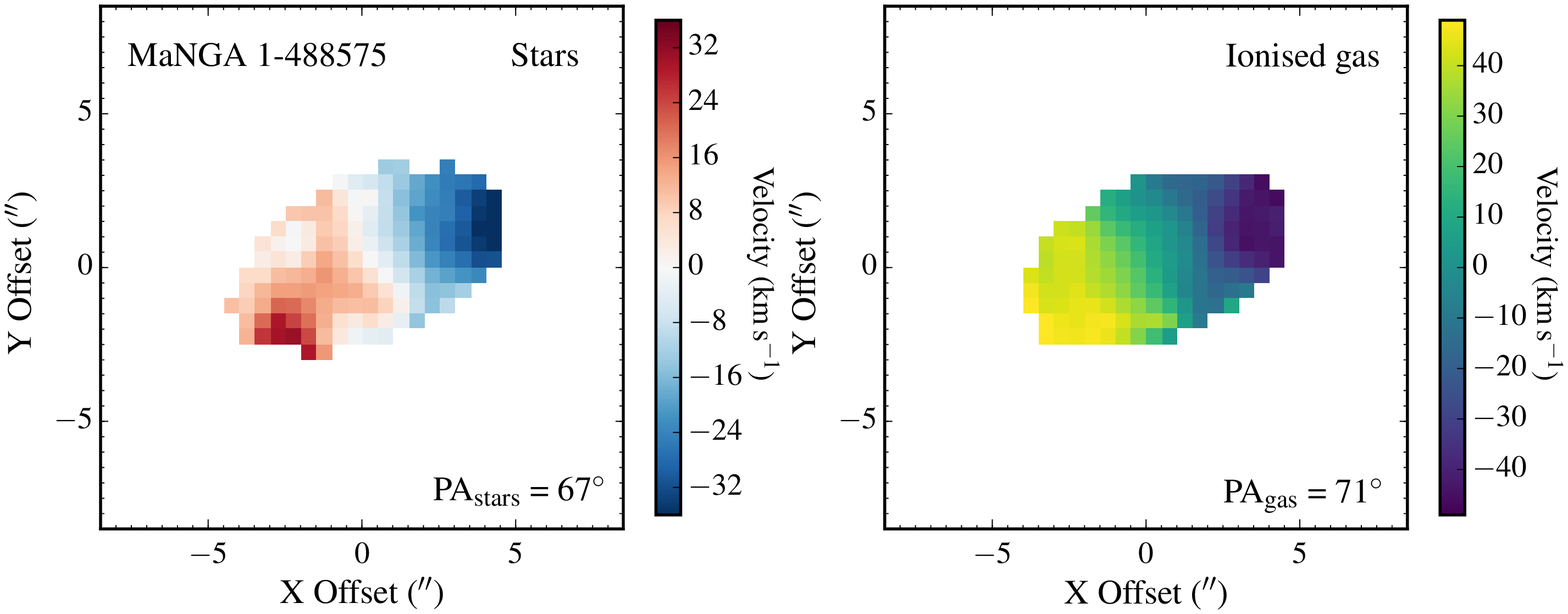}\\
\includegraphics[width=0.8\textwidth]{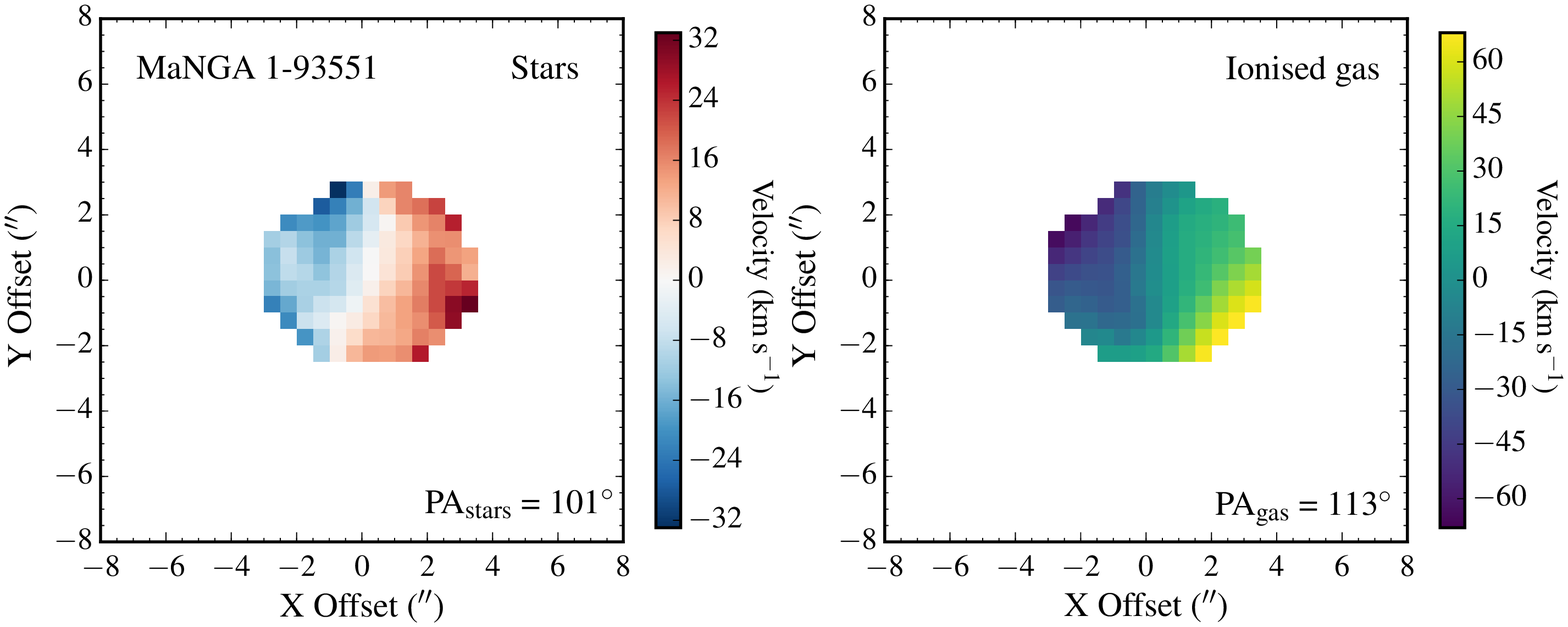}\\
\caption{Stellar velocity maps (left panel) and ionised gas velocity maps (right panel) for the low-mass quenched MaNGA galaxies which contain an ionised gas component. The ionised gas velocity fields are traced using the H$\alpha$ line. The maps for the remaining galaxies are shown in Fig.\ref{fig:vfields}.}
\label{fig:appvfields}
\end{figure*}

\begin{figure*}
\centering
\includegraphics[width=0.8\textwidth]{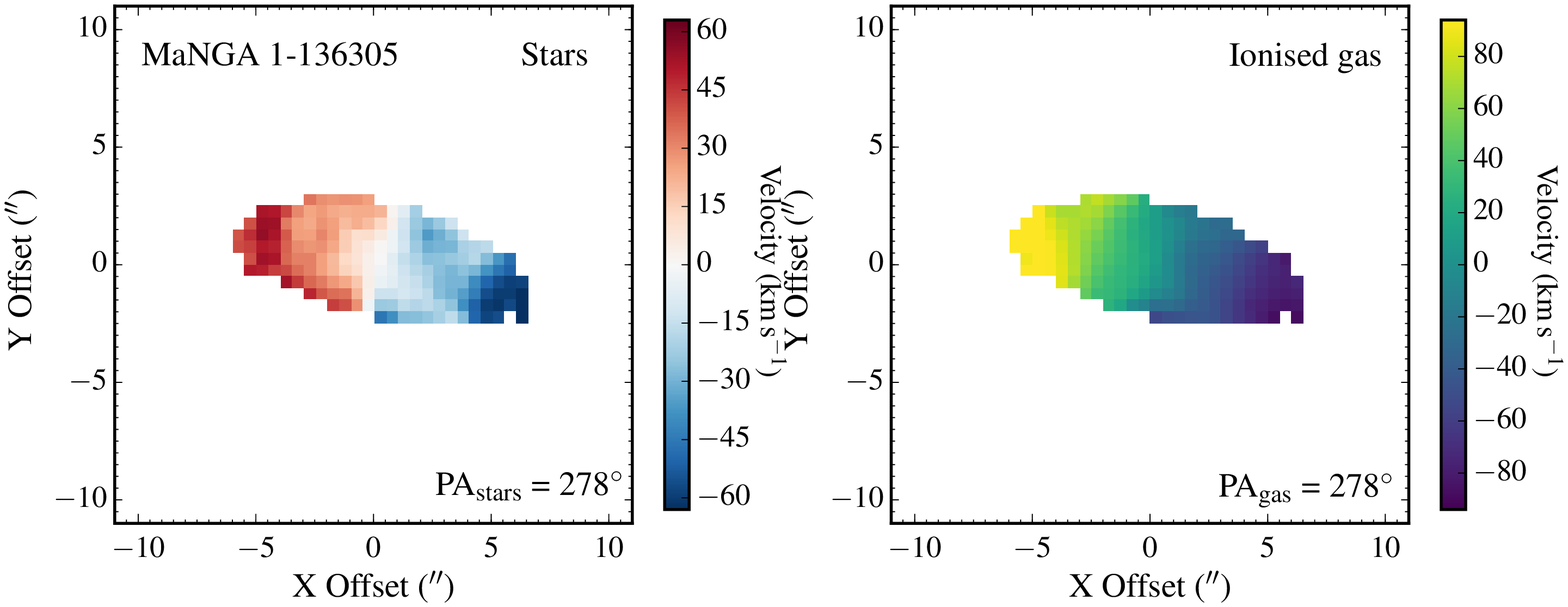}\\
\includegraphics[width=0.8\textwidth]{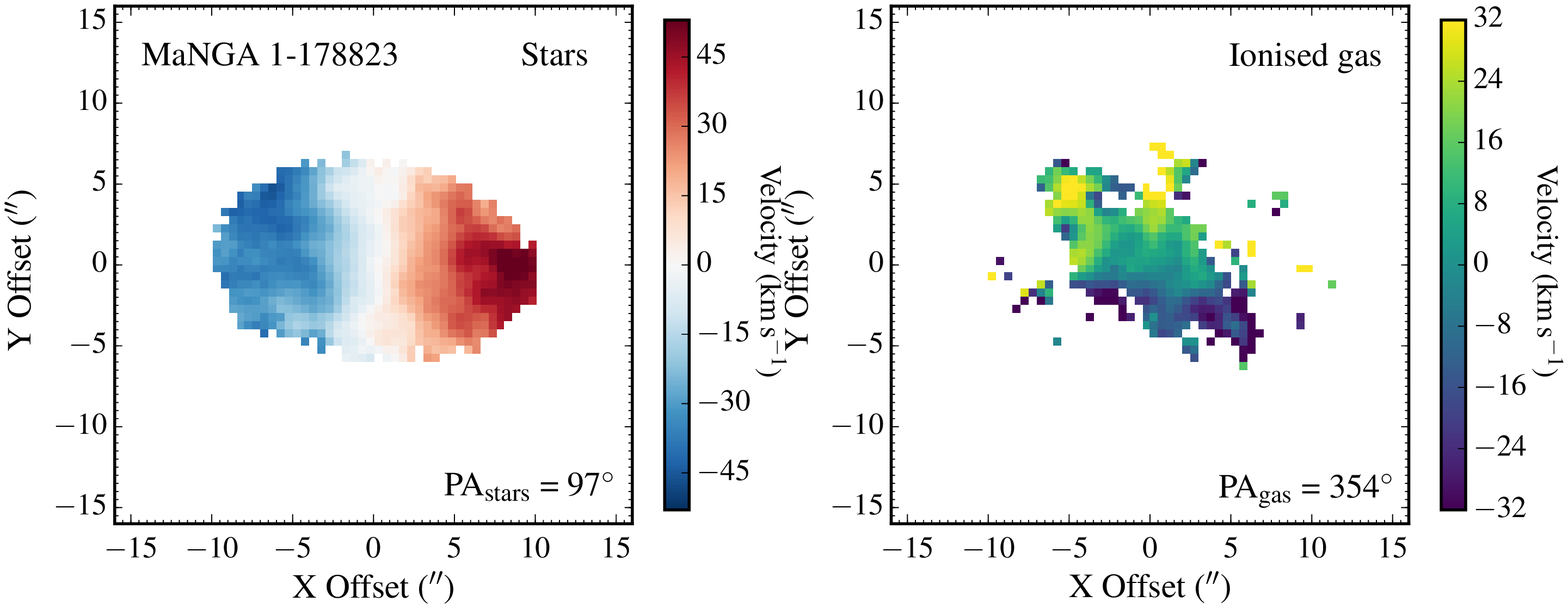}\\
\includegraphics[width=0.8\textwidth]{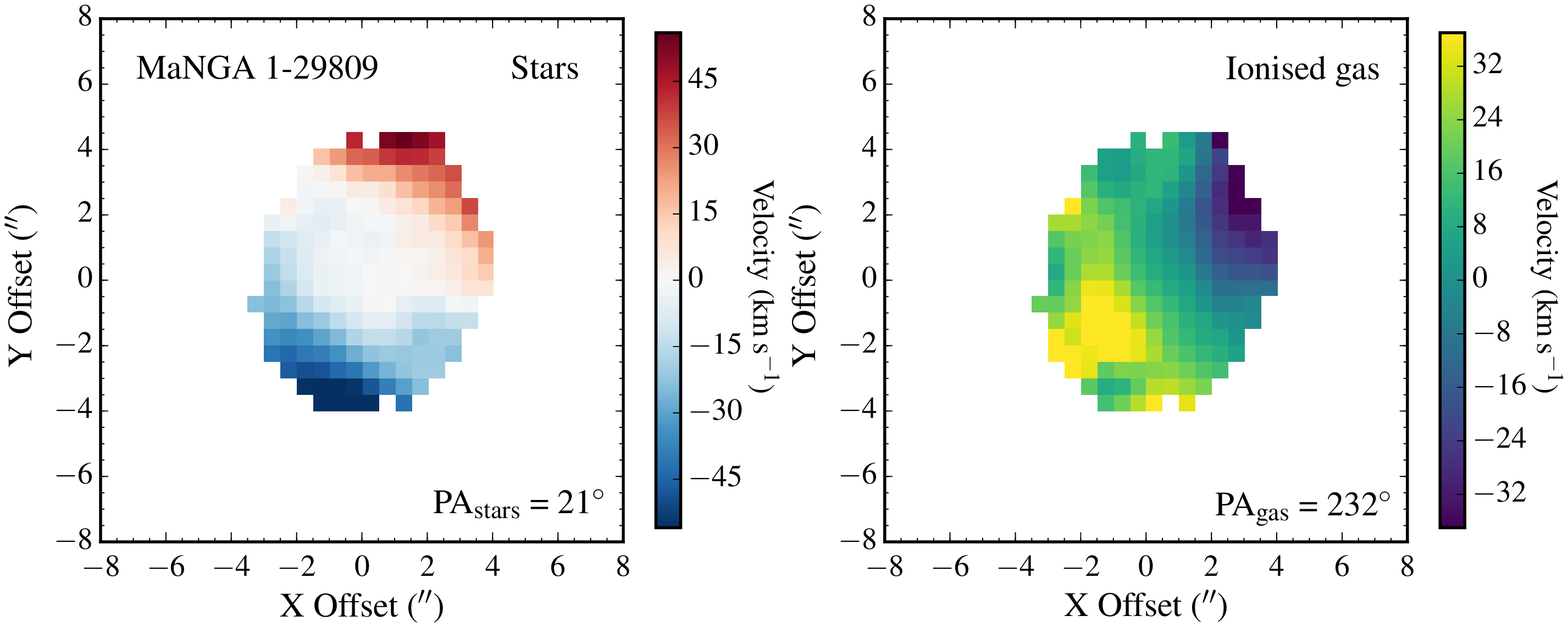}\\
\includegraphics[width=0.8\textwidth]{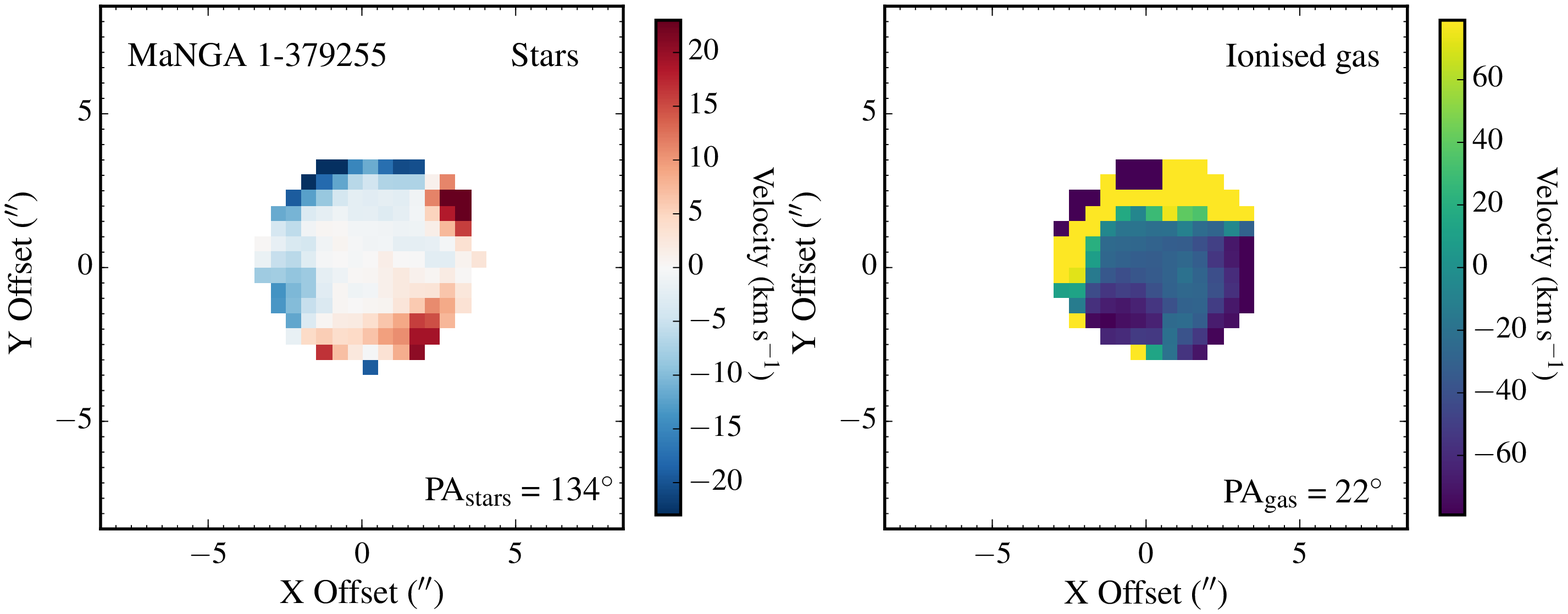}\\
\contcaption{Stellar velocity maps (left panel) and ionised gas velocity maps (right panel) for the low-mass MaNGA galaxies in our sample which contain an ionised gas component. }
\end{figure*}

\begin{figure*}
\centering
\includegraphics[width=0.8\textwidth]{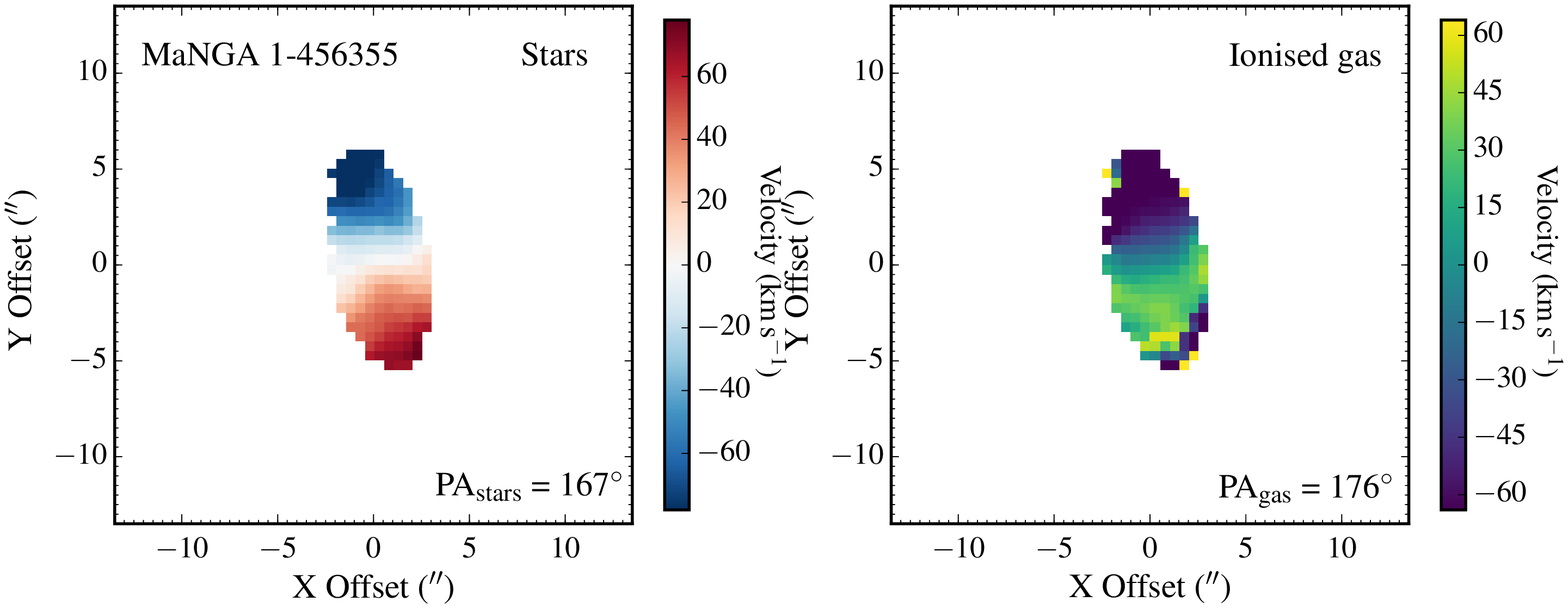}\\
\includegraphics[width=0.8\textwidth]{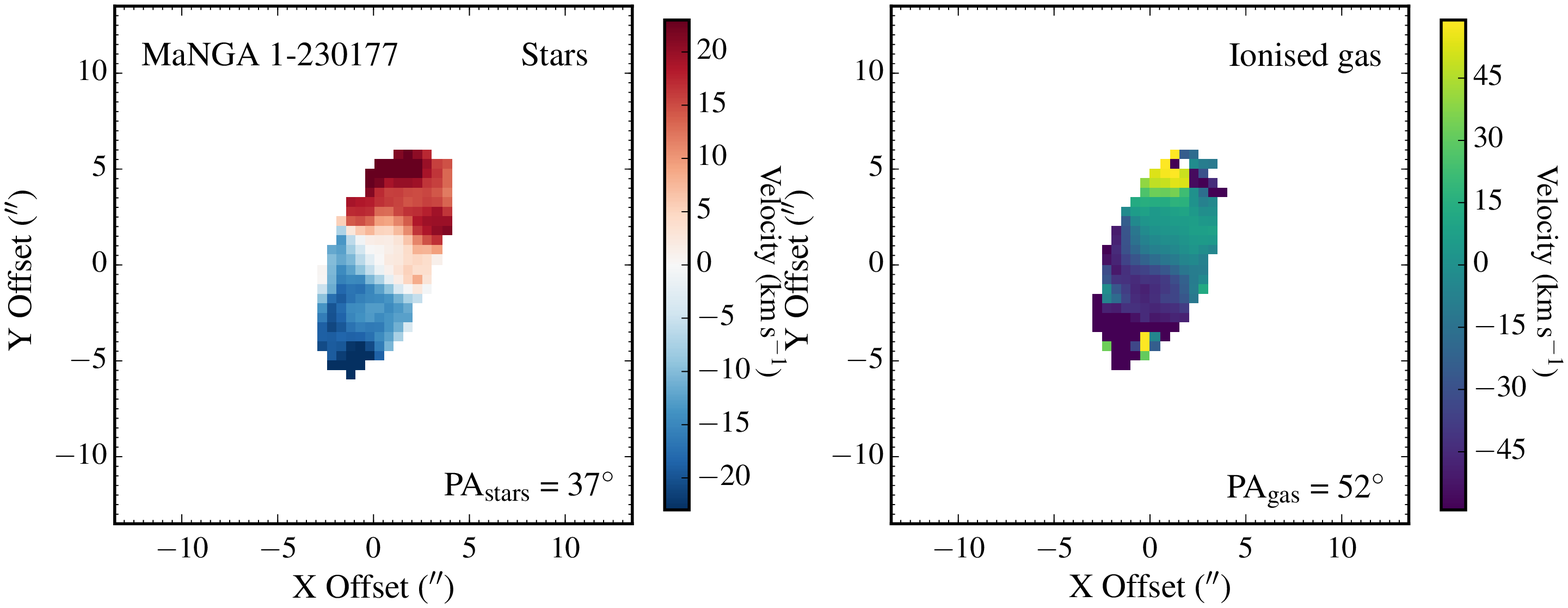}\\
\contcaption{Stellar velocity maps (left panel) and ionised gas velocity maps (right panel) for the low-mass MaNGA galaxies in our sample which contain an ionised gas component.}
\end{figure*}

\section{BPT diagrams for galaxies with kinematically-offset ionised gas}
\label{appendix:bptoff}

\begin{figure*}
 \centering
 \includegraphics[width=0.42\textwidth]{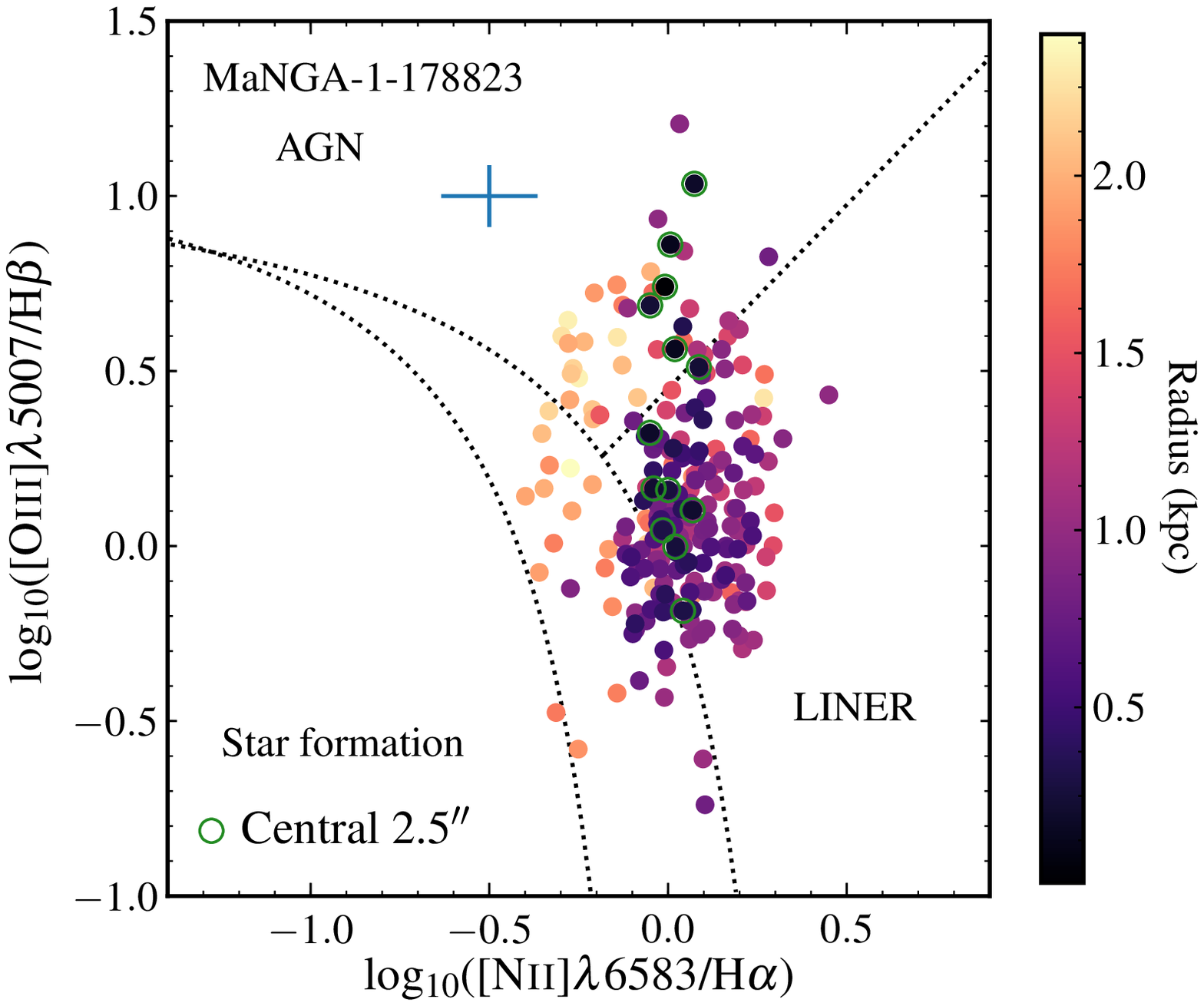}
 \includegraphics[width=0.42\textwidth]{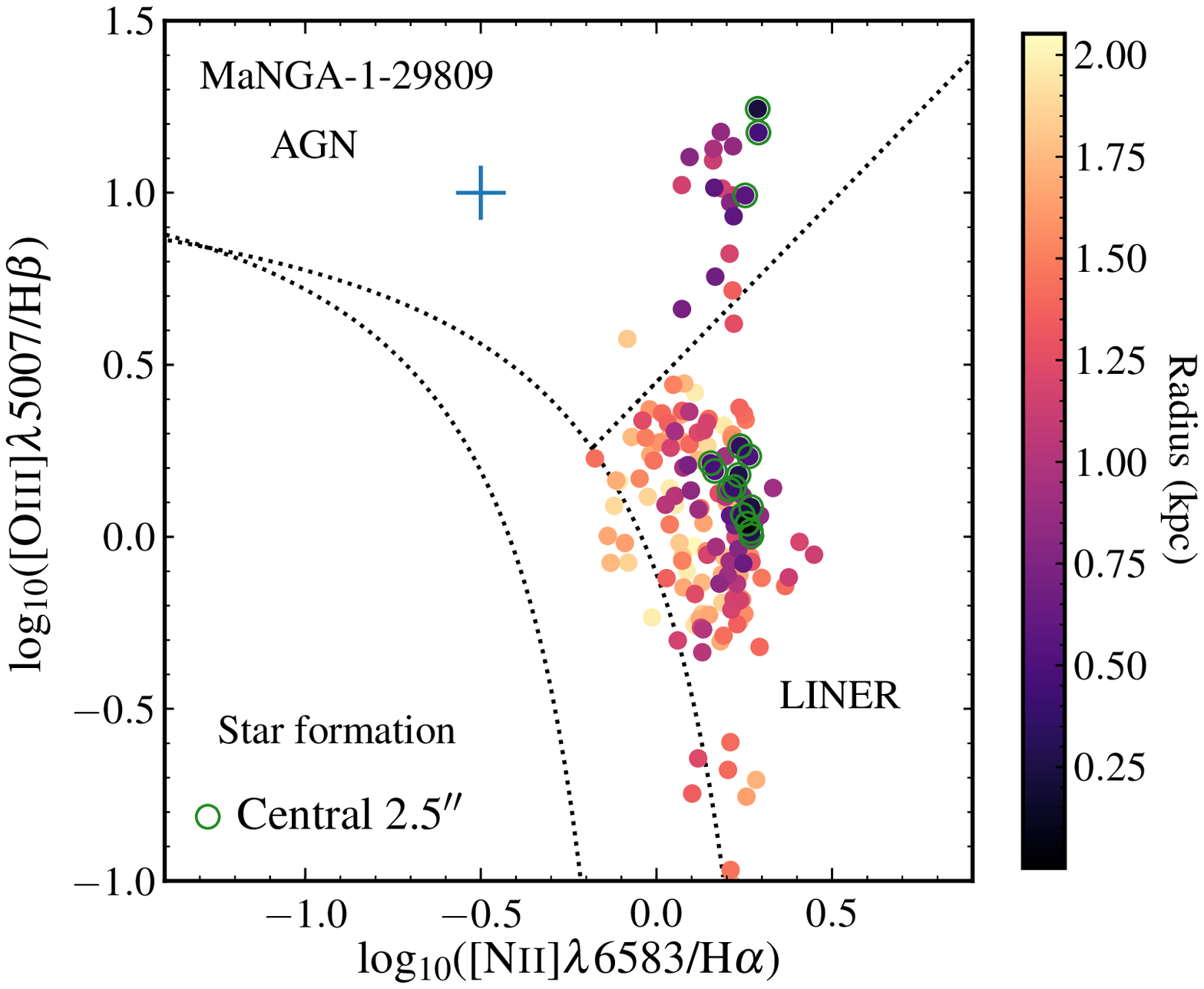}\\
 \includegraphics[width=0.42\textwidth]{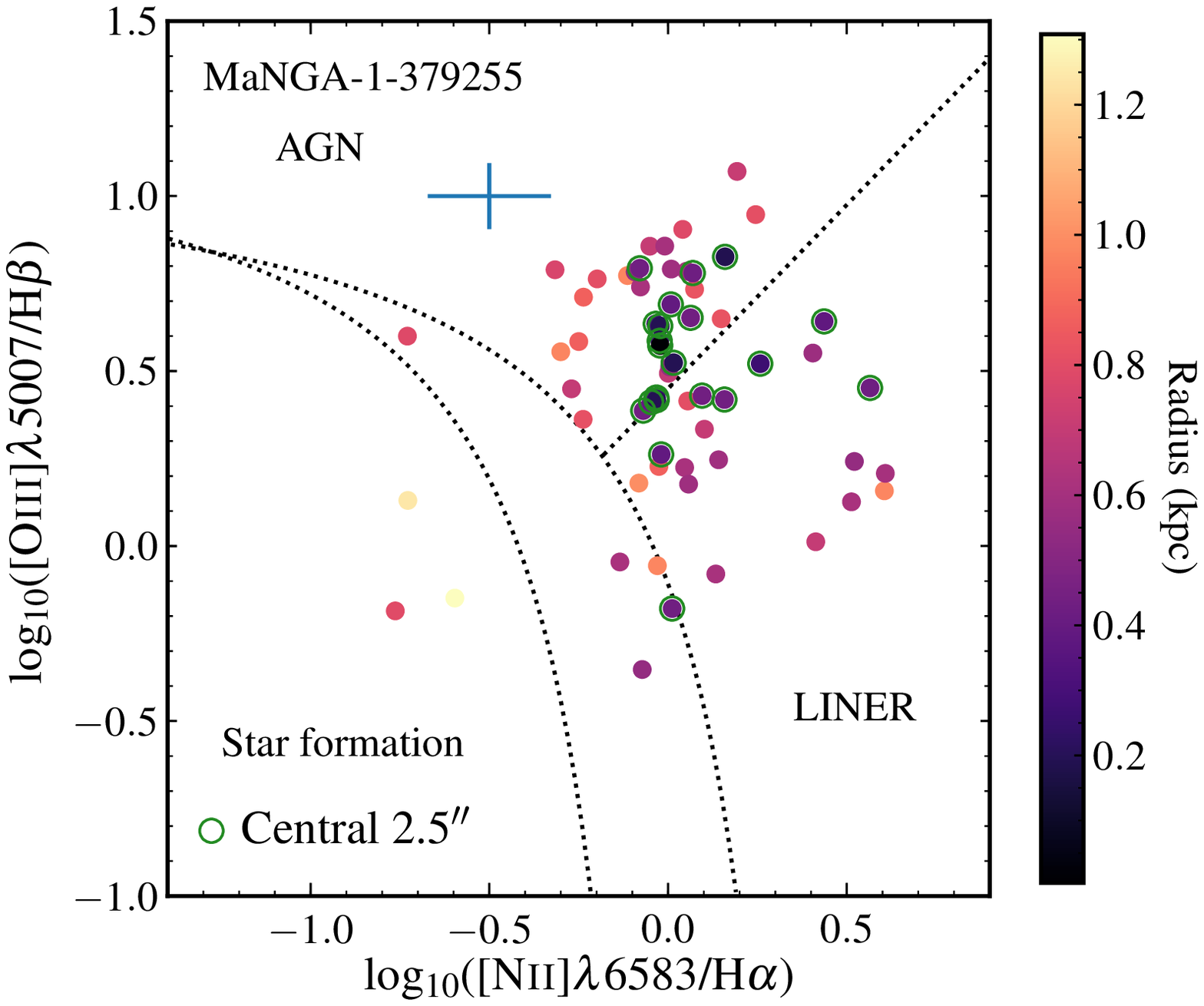}
 \caption{Spaxel-by-spaxel BPT diagrams for the low-mass galaxies with kinematically-offset ionised gas. Spaxels in the central 2.5\,arcsecs (the PSF of the reduced datacubes) are circled in green. Also plotted for diagnostic purposes are the \citet{2001ApJS..132...37K} and  \citet{2003MNRAS.346.1055K} classification lines, which are used to separate extreme starbursts and \ion{H}{ii} regions from AGN-like emission. The \citet{2007MNRAS.382.1415S} division between Seyfert-like and LINER-like (Low Ionisation Nuclear Emission Region) emission line ratios is also plotted. Only spaxels with S/N $>5$ in the H$\alpha$ emission line are plotted.  No object in this sample exhibits central line ratios consistent with star formation. }
\label{fig:appbptoff}
\end{figure*}

Spaxel-by-spaxel BPT diagrams are presented in Fig.\,\ref{fig:appbptoff} for the low-mass galaxies in the sample defined in Sec.\,\ref{sec:whanclass} and Sec.\ref{sec:iongas} that host a kinematically-offset ionised gas component. These galaxies typically have AGN-like emission line ratios at their centres. 

\section{BPT diagrams for galaxies with co-rotating ionised gas}
\label{appendix:bptco}

Spaxel-by-spaxel BPT diagrams are presented in Fig.\,\ref{fig:appbptco} for the low-mass galaxies in the sample defined in Sec.\,\ref{sec:whanclass} and Sec.\ref{sec:iongas} that host an ionised gas component that is either co-rotating or counter-rotating with their stellar component. The majority of these galaxies host composite star forming / AGN-like emission line ratios at their centres, and cannot therefore be classed as AGN-host galaxies. Only spaxels with S/N $>5$ in the H$\alpha$ emission line are plotted.   

\begin{figure*}
\centering
\includegraphics[width=0.42\textwidth]{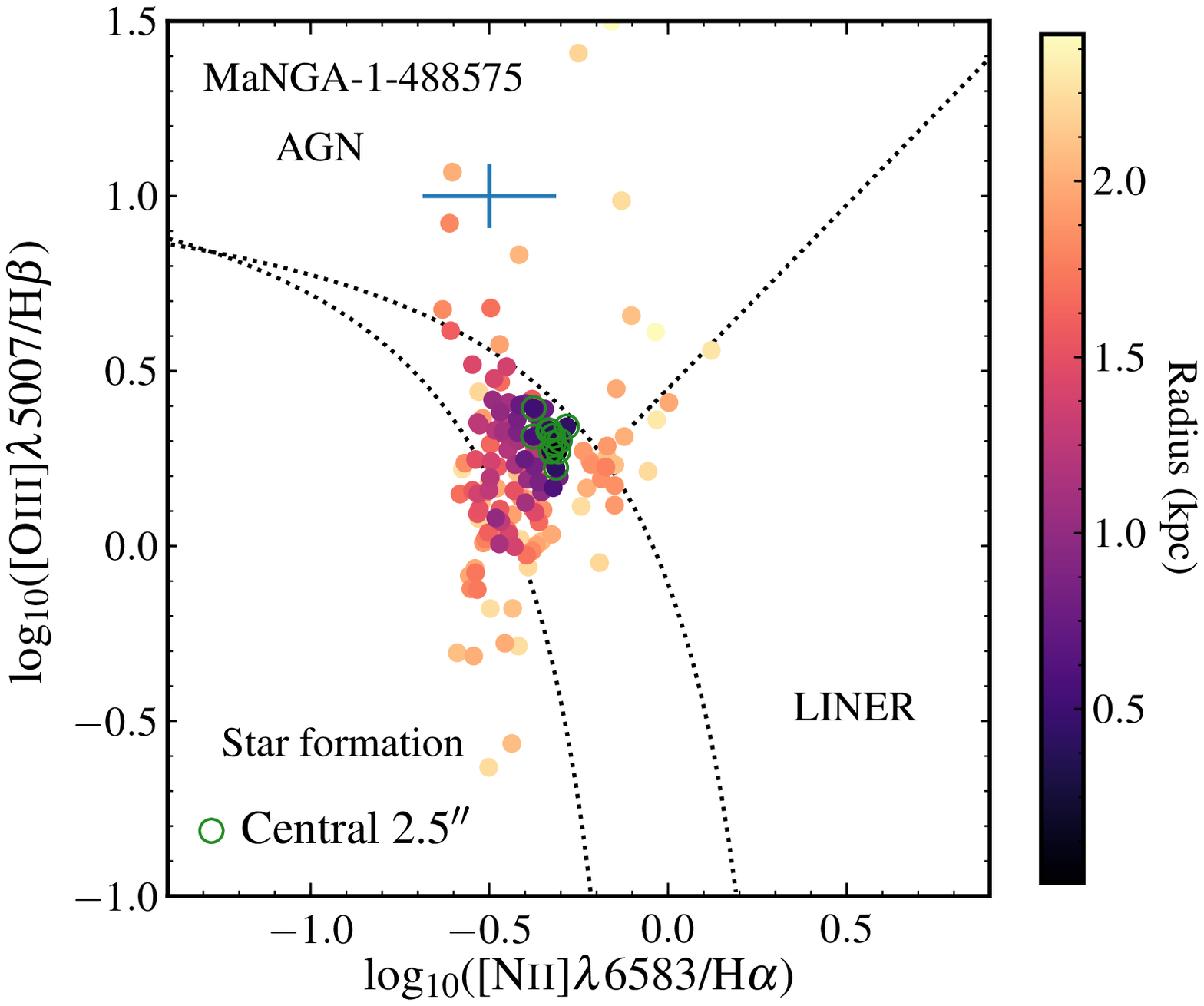}
\includegraphics[width=0.42\textwidth]{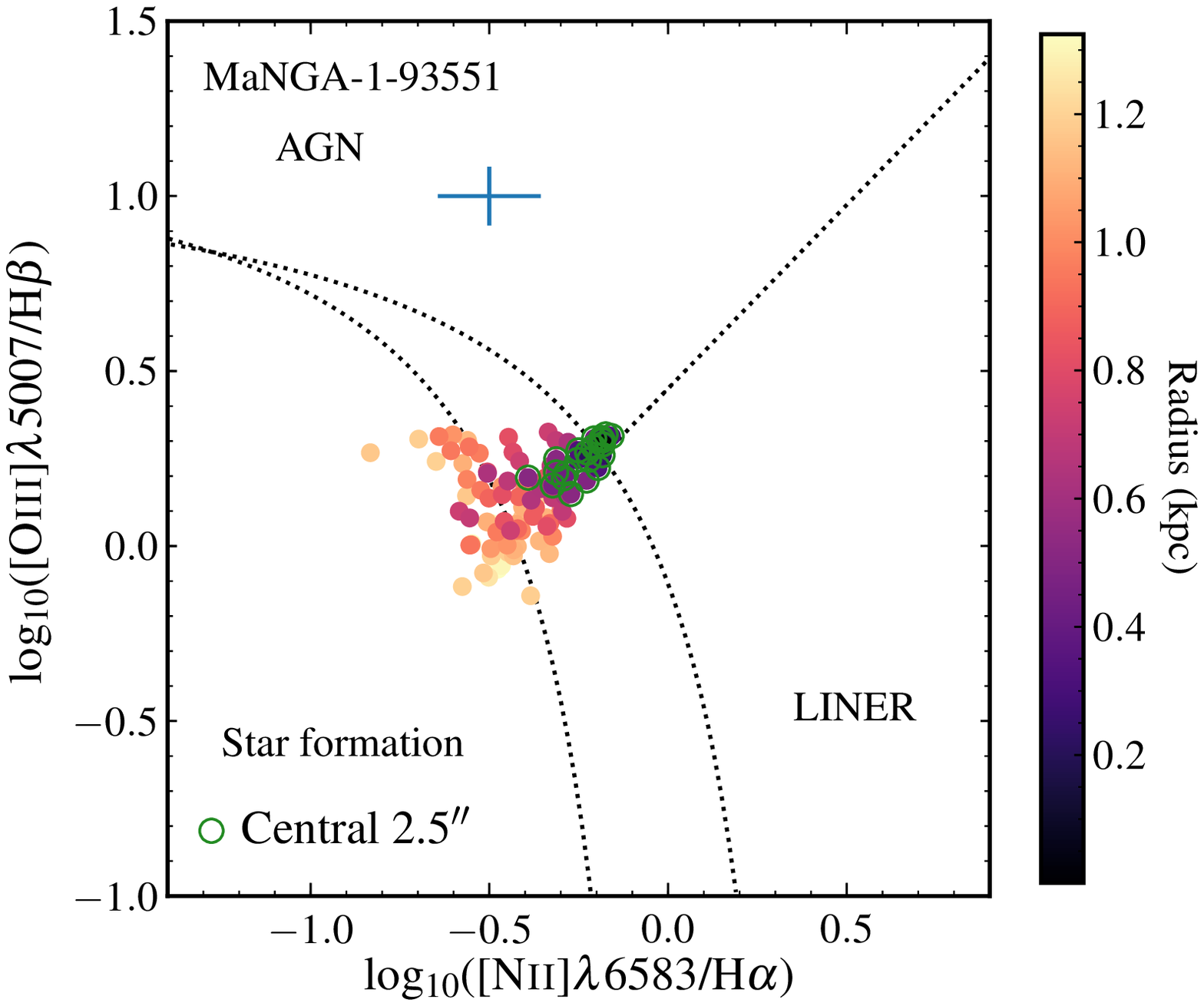}
 \includegraphics[width=0.42\textwidth]{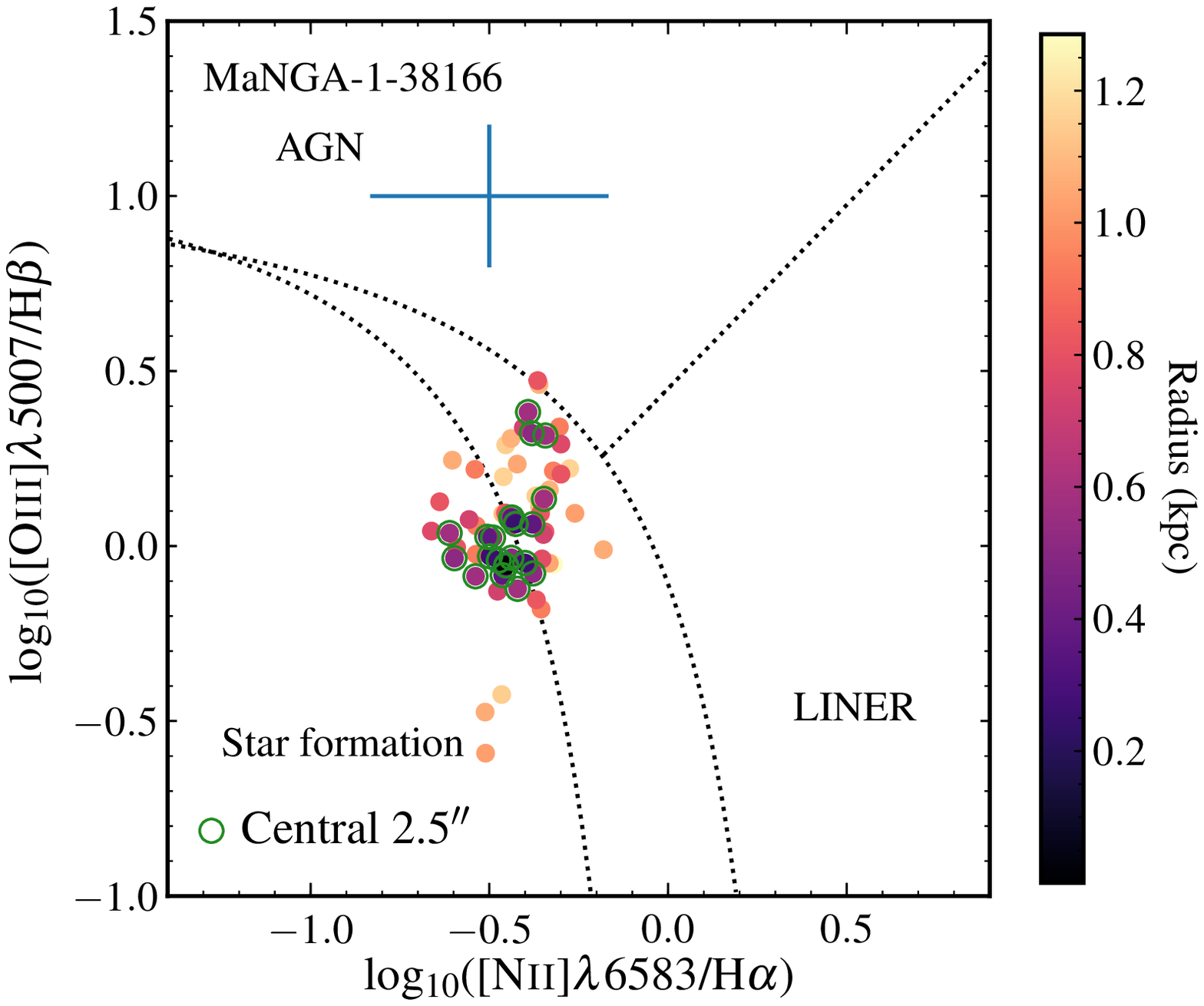}
 \includegraphics[width=0.42\textwidth]{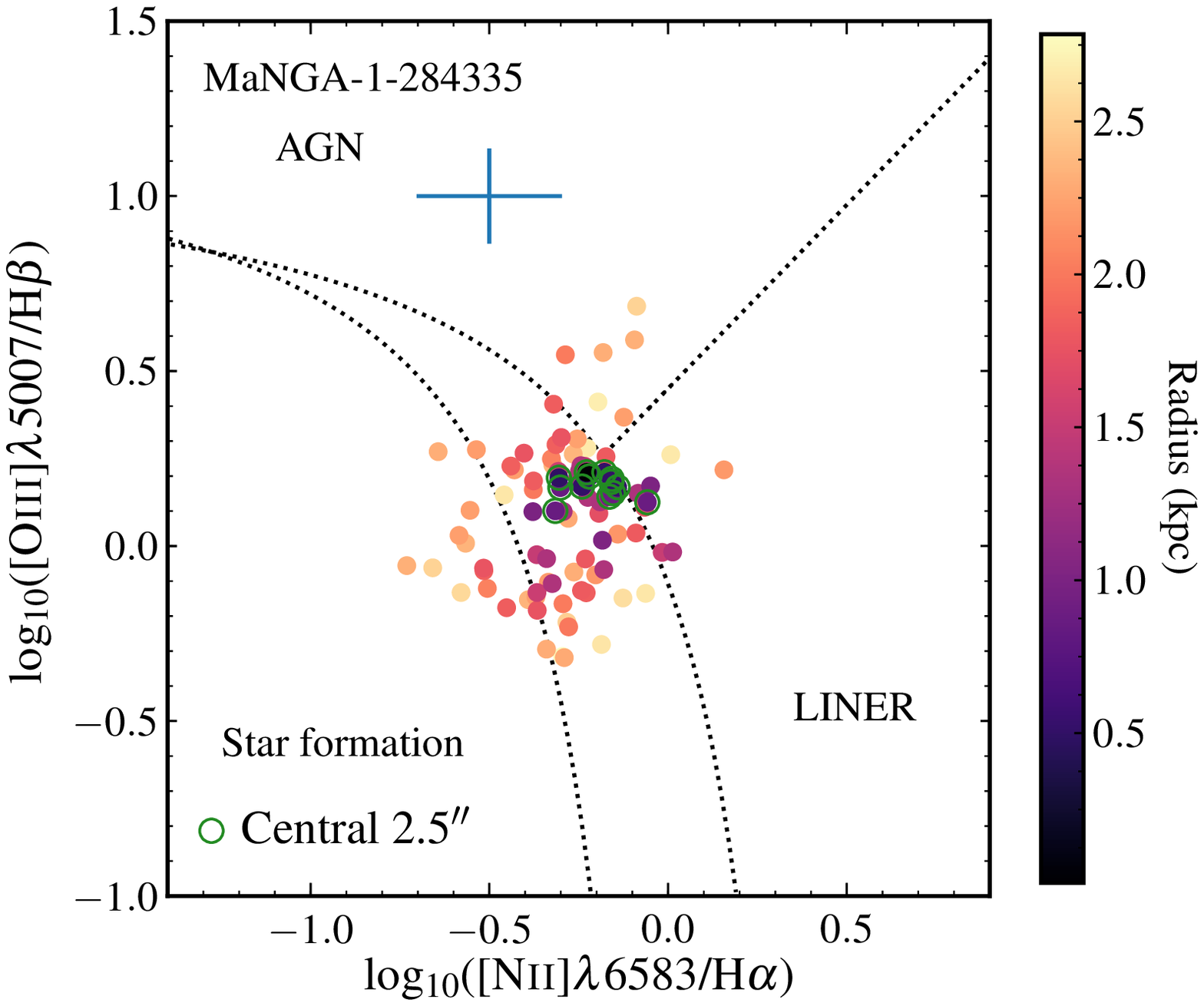}\\
\caption{Spaxel-by-spaxel BPT diagrams for the galaxies in our sample with gas that is either co- or counter-rotating from their stellar component. The colour of each point corresponds to its separation from the centre of the galaxy in arcseconds. Spaxels in the central 2.5\,arcsecs (the PSF of the reduced datacubes) are circled in green. Only spaxels with S/N $>5$ in the H$\alpha$ emission line are plotted.   
MaNGA-1-230177 has central spaxels that exhibit emission line ratios consistent with AGN activity. } 
\label{fig:appbptco}
\end{figure*}

\begin{figure*}
\centering
 \includegraphics[width=0.42\textwidth]{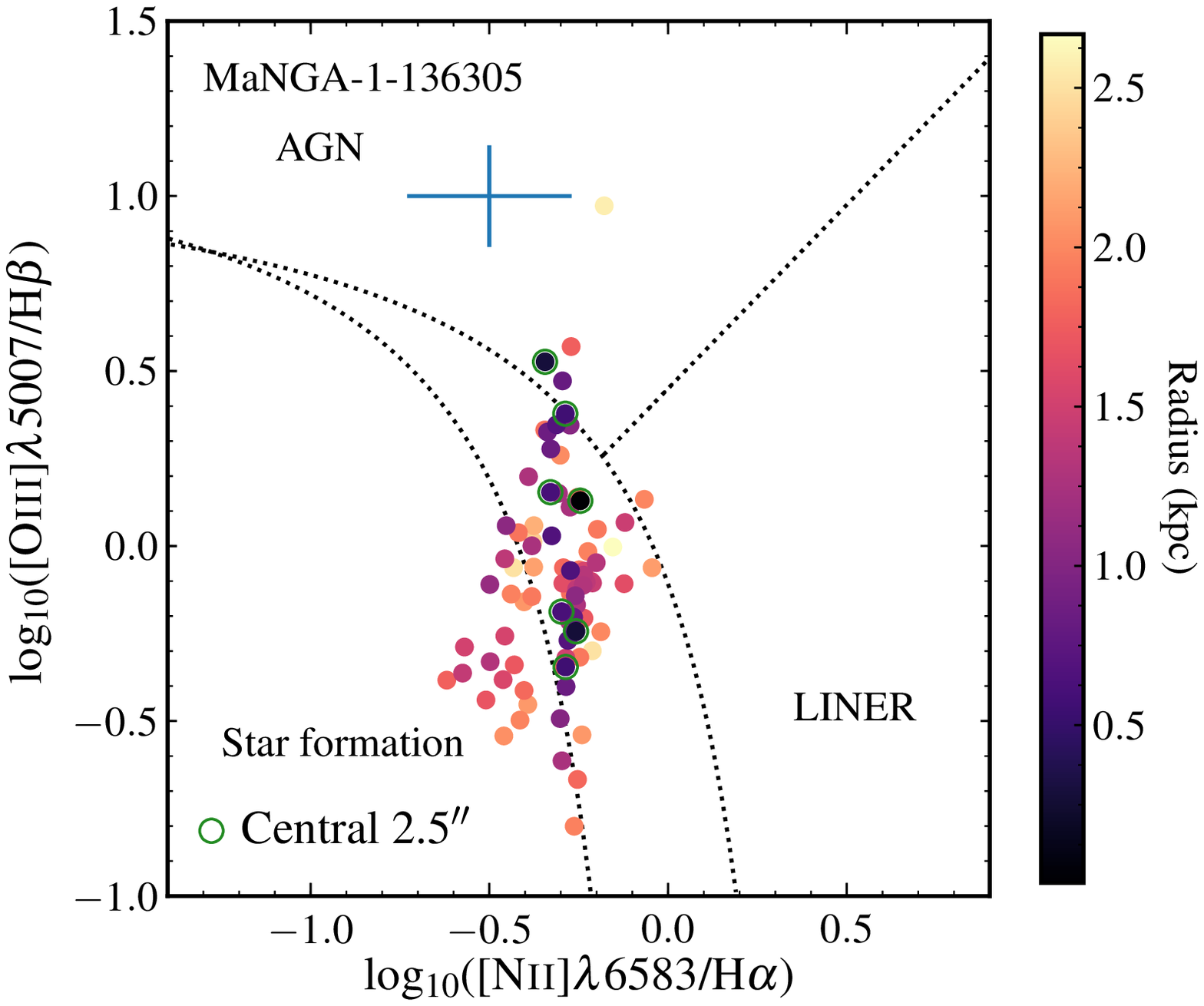}
 \includegraphics[width=0.42\textwidth]{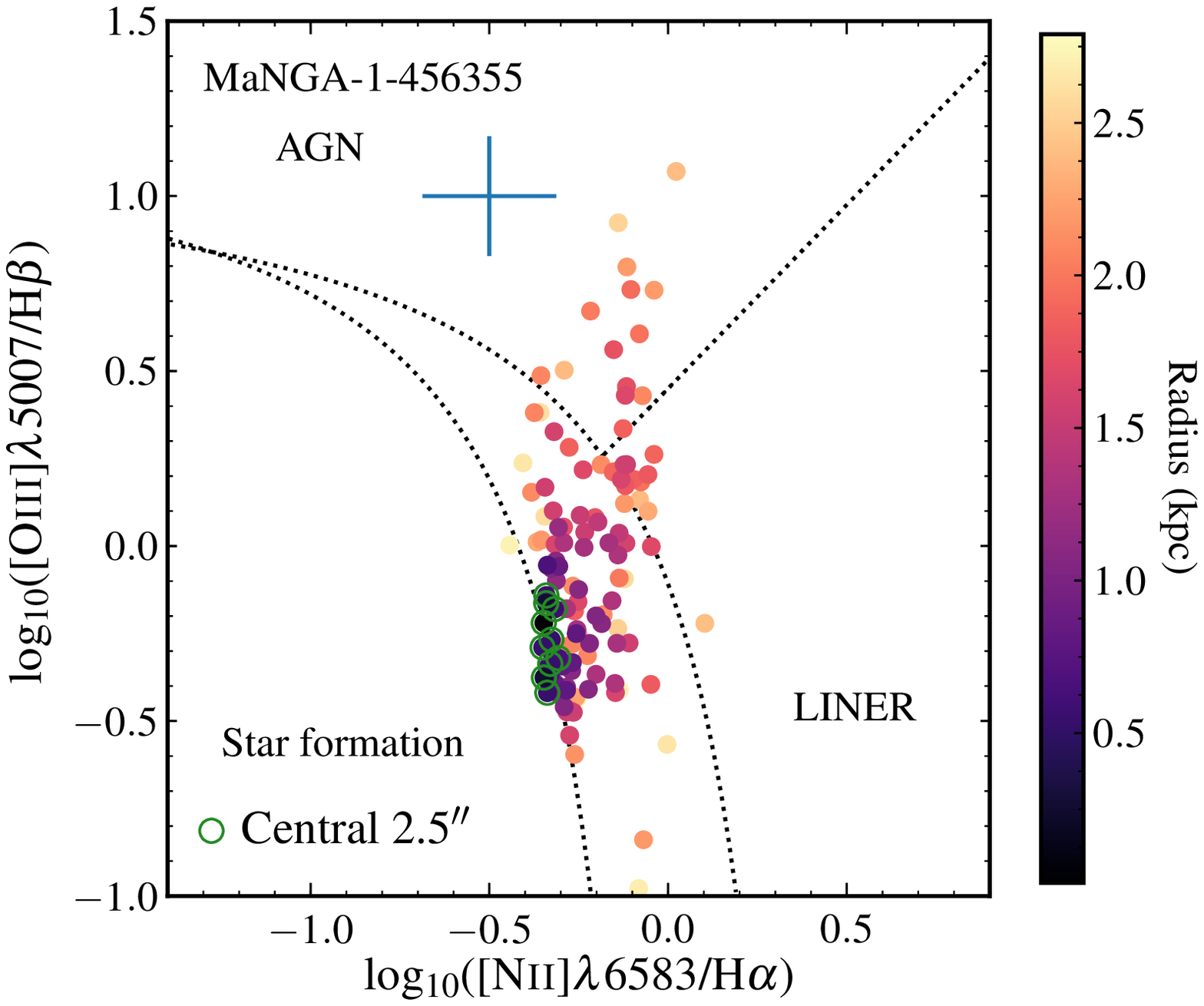}\\
 \includegraphics[width=0.42\textwidth]{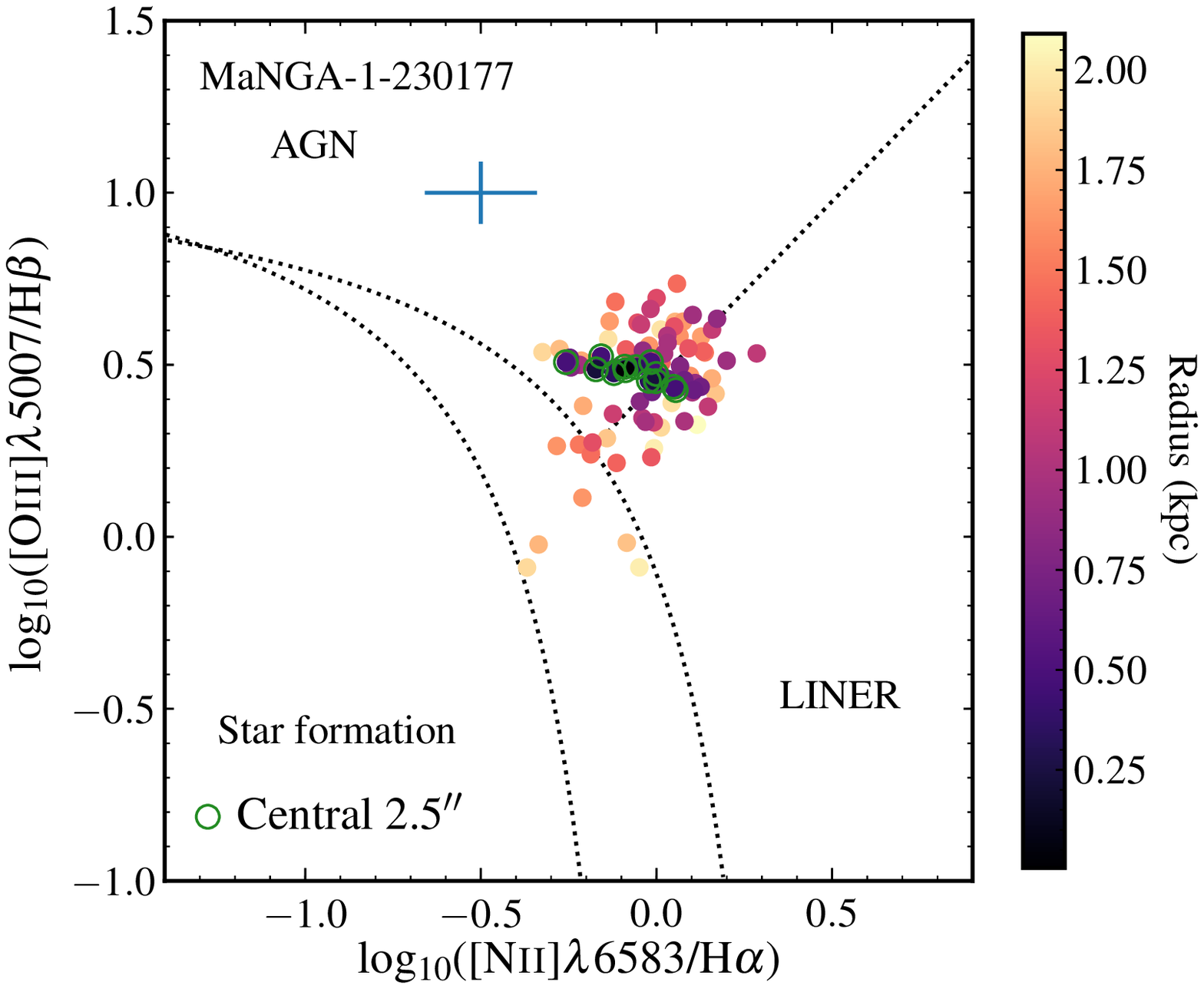}
 \contcaption{Spaxel-by-spaxel BPT diagrams for low-mass, quenched MaNGA galaxies which host an ionised gas that is either co- or counter-rotating with their stellar component. MaNGA-1-230177 has central spaxels that exhibit emission line ratios consistent with AGN activity. }
\end{figure*}


\bsp	
\label{lastpage}
\end{document}